# Size measurements and characterization of the astrophysical properties of multiple-component radio-AGNs in the ROGUE I catalog

Arti Goyal,[1] Arpita Misra,[1] Subhrata Dey,[2] Unnikrishnan Sureshkumar,[3,2] Marian Soida,[1] Anna Wójtowicz,[4] Grażyna Stasińska,[5] Natalia Vale Asari,[6] and Syed Naqvi[7,8]

[1]Astronomical Observatory of the Jagiellonian University, ul. Orla 171, 30-244 Kraków, Poland
[2]National Center for Nuclear Research, ul. Pasteura 7, 02-093 Warsaw, Poland
[3]Wits Centre for Astrophysics, School of Physics, University of the Witwatersrand, Private Bag 3, Johannesburg 2050, South Africa
[4]Department of Theoretical Physics and Astrophysics, Faculty of Science, Masaryk University, Kotlářská 2, Brno 61137, Czech Republic
[5]LUX, Observatoire de Paris, Université PSL, Sorbonne Université, CNRS, 92190 Meudon, France
[6]Departamento de Fìsica–CFM, Universidade Federal de Santa Catarina, C.P. 5064, 88035-972, Florianópolis, SC, Brazil
[7]Institute of Nuclear Physics, Polish Academy of Sciences, Radzikowskiego 152, 31-342 Kraków, Poland
[8]Department of Physics, Indian Institute of Technology Madras, Chennai 600036, India

## ABSTRACT

We present hand-curated size measurements for a sample of 2,002 multiple-component radio AGNs in the Radio sources associated with Optical Galaxies and having Unresolved or Extended morphologies I (ROGUE I) catalog. The sources span total angular sizes of $\sim 5''$–$1{,}100''$ which translates to projected linear sizes $\sim$10 kpc– $\sim$2 Mpc across $0.01 \leq z \leq 0.54$. About 10% of the sample are compact ($\leq$60 kpc) while $\sim$3% are giant radio sources ($\geq$700 kpc). Roughly 34% are associated with galaxy clusters, and 16% exhibit an arm-length asymmetry ratio $\geq$2. The cluster association fractions are comparable across Fanaroff-Riley (FR) I, II, and I/II type morphological classes. Arm-length asymmetries occur at similar rates in FR I and I/II classes but are about a factor of 2–4 less common in FR II, supporting the view that their jets are more stable and collimated, and thus less prone to disruption in dense cluster environments. In contrast, bent-angle sources (wide- and narrow-angle tails) show arm-length asymmetries about a factor of four less frequently than cluster-associated sources, suggesting that asymmetries are smoothed out by the local intracluster medium conditions. The mean galaxy number densities of FR I, II, I/II, wide-angle, narrow-angle, and head-tail sources are comparable, supporting the dominant role of local intra-cluster medium conditions in shaping bent morphologies. Radio power–linear size tracks for FR IIs show that the population is dominated by AGNs with jet kinetic powers $\leq 10^{38}$ erg s$^{-1}$ and comprising both young ($\leq$10 Myr) and old ($\sim$100 Myr) AGNs. We also compare z, angular/linear sizes, core/total radio luminosities, optical magnitudes, black hole, and stellar masses across the morphological classes.

## 1. INTRODUCTION

It is remarkable that a relatively simple progenitor, that is, a supermassive black hole (SMBH), when accreting matter from its surroundings and sometimes ejecting bipolar, relativistic, and magnetized plasma outflows can manifest itself in a complex phenomenon of jetted active galactic nuclei (AGN; M. C. Begelman et al. 1984; R. Blandford et al. 2019), comprising a wide range of radio powers, morphologies, and linear sizes (see Figure 2 of M. J. Hardcastle & J. H. Croston 2020). The combination of different properties such as the accretion flow (radiatively efficient vs. radiatively inefficient accretion flows; P. N. Best & T. M. Heckman 2012; B. Mingo et al. 2014), intrinsic jet power *vis-à-vis* optical luminosity of the host (B. L. Fanaroff & J. M. Riley 1974; M. J. Ledlow & F. N. Owen 1996), a profile of the external medium density (C. R. Kaiser & P. Alexander 1997;

Email: arti.goyal@uj.edu.pl

K. M. Blundell et al. 1999; C. R. Kaiser & P. N. Best 2007), and the jet composition (electron-position pair plasma vs. electron-proton plasma C. S. Reynolds et al. 1996; A. Celotti et al. 1997; J. H. Croston et al. 2018) appear to be the deciding factors for Fanaroff-Riley (FR) type I and II (B. L. Fanaroff & J. M. Riley 1974) morphologies of radio galaxies. These, as well as some other physical properties, can shape the resulting morphology (e.g., J. H. Croston et al. 2018; B. Mingo et al. 2019). Besides the traditional classification of radio AGNs into FR type-I where jets' brightness decreases monotonically with distance into diffuse lobes and plums on kpc scales, and FR type-II, where well-collimated jets terminate into high surface brightness hot-spots morphologies, a significant fraction also exhibit a hybrid FR morphology where one lobe has type-I morphology while the counter-lobe has type-II morphology; (I/II; Gopal-Krishna & P. J. Wiita 2000; A. D. Kapińska et al. 2017; S. Kumari & S. Pal 2021). The FR types I, II, and I/II morphologies are further classified as 'straight-line FR' sources, as in these galaxies, the jets/hot spots/lobes and the host galaxy can be connected by a straight line. We also find radio AGNs with Wide-angle tail (WAT) and Narrow-angle tail (NAT) morphologies, where jets/lobes lie on one side of the radio core and have lobe opening angles >90° and <90°, respectively (F. N. Owen & L. Rudnick 1976; C. P. O'Dea & F. N. Owen 1985; V. Missaglia et al. 2019; S. Pal & S. Kumari 2021). Moreover, Head-tail (HT) sources have only one set of jet/lobe (L. Rudnick & F. N. Owen 1976; B. S. Koribalski et al. 2024). WAT, NAT, and HT sources are classified as 'bent-angle' sources and provide crucial information on the environment in which they reside (C. P. O'Dea & S. A. Baum 2023, for a recent review). Double-double (DD) radio AGNs have two sets of distinct lobes on either side of the radio core. These sources exhibit recurrent AGN activity and are crucial for estimating the AGN duty cycle (D. J. Saikia et al. 2006; C. Konar et al. 2013; V. H. Mahatma et al. 2019). X-shaped radio AGNs display two pairs of jets/lobes at right angles to each other from the radio core (C. C. Cheung 2007; W. D. Cotton et al. 2020). The X-shaped radio galaxies can form as a result of spin-flip of the SMBH or due to the deflection of the backflowing jet plasma by the thermal haloes of the host galaxy ( Gopal-Krishna et al. 2003; L. Saripalli & R. Subrahmanyan 2009). Objects where the lobe and counter lobe connect to the optical host/radio core in a Z-shape fashion are classified as Z- or S- shaped. The curved morphology of the radio-emitting plasma is possibly caused by the precession of the jet (S. Sethi et al. 2024; A. Misra et al. 2025).

A crucial parameter related to the evolution of radio AGNs, is the estimation of its total angular size (or projected total linear size), which holds significance for a variety of reasons: (1) historically, the evolution of angular size ($\theta$) with redshift and flux density has been used to derive constraints on cosmological models (K. I. Kellermann 1972; V. K. Kapahi 1975, 1989) and investigate whether there is intrinsic evolution within the population (A. K. Singal 1988; K. M. Blundell et al. 1999), (2) gives a proxy for source age (S. A. E. G. Falle 1991), (3) gives the age distribution and kinetic jet power (M. J. Hardcastle 2018), (4) helps study the cluster environment (J. H. Croston et al. 2019; E. Vardoulaki et al. 2021). In addition, modeling of AGN evolution has been challenging due to the complex relationship between the jet kinetic power and density profile in which they grow, even when considering FR type-I and type-II classes (M. J. Hardcastle 2018; R. J. Turner et al. 2023). Furthermore, as radio AGNs exhibit a variety of radio morphologies, any modeling must self-consistently reproduce all these shapes, motivating us to identify distinguishing features among the radio galaxy population that could serve as testbeds for the physical models frequently employed.

The biggest challenge in comparative studies is to obtain homogeneously selected samples of different AGN classes with a *sufficient* number of sources to obtain a meaningful statistic. Such samples, although flux-limited, will, in principle, be free of other selection biases (i.e., AGN samples made at different observing frequencies) and, as such, can provide an unbiased view of the physical picture. The Radio sources associated with Optical Galaxies and having Unresolved or Extended morphologies I (ROGUE I; D. Kozieł-Wierzbowska et al. 2020) catalog provides such AGN samples. It gives a large number of sources (AGNs and star-forming galaxies) that were found to be associated with single-component (one radio component associated with the optical host galaxy) and multiple-components (i.e., having two or more radio components associated with the optical host galaxy) at the central radio frequency of 1.4 GHz. In the category of multiple-component radio AGN classes, we include I, II, I/II, DD, X, Z, WAT, NAT, and HT type radio AGNs.

The present study measures the total angular/linear sizes and compares the physical properties of multiple-component radio AGNs across the AGN subclasses. Throughout this paper, we use the concordant cosmology with the Hubble constant $H_0 = 69.6$ km/s/Mpc, $\Omega_{\rm M} = 0.286$, and $\Omega_\Lambda = 0.714$ (C. L. Bennett et al. 2014), according to that adopted in D. Kozieł-Wierzbowska et al. (2020). The paper is organized as follows. Section 2 gives a brief description of the ROGUE I catalog,

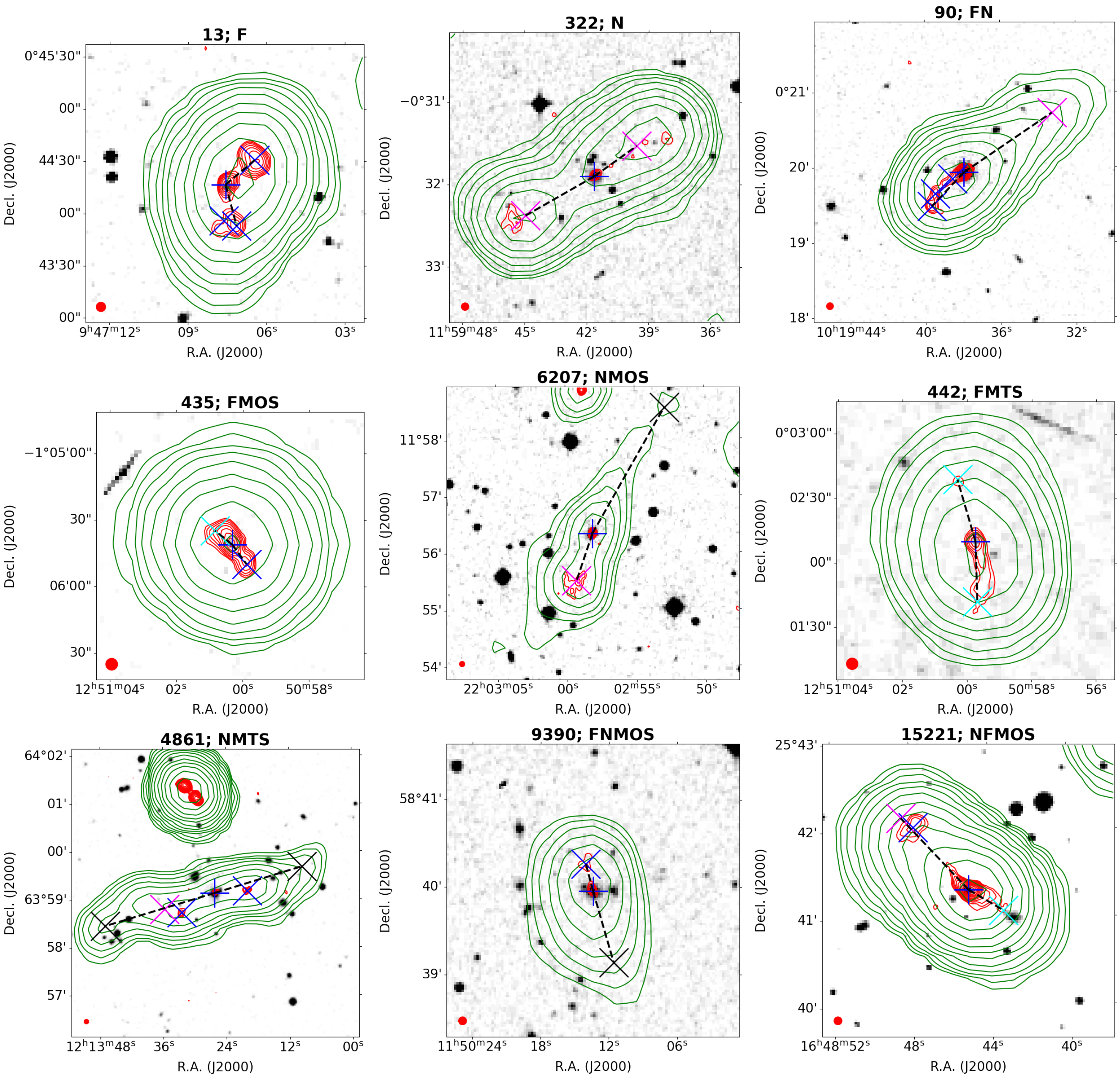

**Figure 1.** Examples of the methodology followed for selecting the farthest lobe and the counter-lobe components for a radio AGN. Each panel presents a radio-optical overlay map of the ROGUE I galaxy, where the FIRST and NVSS intensities are shown by red and green contours, respectively, while the DSS intensity is shown in grayscale. The radio intensities are drawn at typical $3\sigma$ sensitivity limits of these surveys ($\sigma$ being 0.2 mJy beam$^{-1}$ and 0.45 mJy beam$^{-1}$ for FIRST and NVSS, respectively) and increase by $(\sqrt{2})^n$ where $n$ range from 0,1,2,..,20. A blue plus sign marks the optical position of the host galaxy from the SDSS survey, while cross symbols mark the positions of the lobe components from the radio FIRST/NVSS surveys. Blue and magenta crosses indicate the positions of components detected in the FIRST and NVSS catalogs, respectively. Cyan and black crosses indicate the position of a radio component measured manually using the DS9 from the FIRST and NVSS maps, respectively. For clarity, the radio components belonging to the FIRST core with separation $\leq 3''$ and the NVSS core with separation $\leq 15''$ from the optical position, are not shown. The dashed black lines connect the optical center to the farthest radio components used for size estimation. The ROGUE I ID of galaxy and the keyword describing the method of size measurement (column 16; Table 1) is given at the top of each panel, while the FIRST beam, shown with a red filled circle, is indicated at the bottom left corner.

Section 3 gives details on linear size measurements and other properties, Section 4 gives the discussion on the obtained results, while Section 5 summarizes our main findings.

## 2. DATA

Covering nearly one-third of the sky, the ROGUE I catalog provides *visual* identifications of the radio and optical morphologies for a sample of 32,616 spectroscopically selected optical galaxies. The ROGUE I sample was selected from a parent sample of nearly 670,000 galaxies within a redshift range of 0.01–0.72 from the Sloan Digital Sky Survey (SDSS) Data Release (DR) 7, those presenting a radio component (dubbed as radio core) from the Faint Images of Radio Sky at Twenty-Centimeters (FIRST; R. H. Becker et al. 1995) catalog within a $\leq 3''$ distance (see D. Kozieł-Wierzbowska et al. 2020, for details on the parent sample). Centered on the positions of optical galaxies, the radio contour maps of angular sizes equal to 1 Mpc diameter (at the source distances) were made from the FIRST and the NRAO VLA Sky Survey (NVSS; J. J. Condon et al. 1998) datasets, starting at nominal $3\sigma$ sensitivities of 0.6 mJy beam$^{-1}$ and 1.35 mJy beam$^{-1}$, respectively, enabling us to classify radio morphologies by combining the excellent point source sensitivity of the FIRST survey ($\sim$1 mJy beam$^{-1}$) and the extended emission sensitivity of the NVSS survey ($\sim$2.5 mJy beam$^{-1}$). We overlaid the radio intensities on optical images from the Digitized Sky Survey (DSS). The optical morphological classification was made using SDSS DR 14 maps. The radio-optical overlays and the SDSS maps used for morphological classification of the ROGUE I galaxies are publicly available[9]. The radio emission for the majority ($\sim$92%) of the galaxies was found to be associated with a single-component morphology, primarily the unresolved/partially resolved radio core. In contrast, a small fraction ($\sim$8%) was found to have a multiple-component morphology composed of a radio core and lobe(s).

In the ROGUE I catalog, the single-component radio sources were assigned to have either compact (C) or elongated (E) morphologies based on whether the deconvolved major and minor axes size of the beam from the FIRST survey was 0 or not, respectively. The radio morphological classification of multiple-component sources was based on their appearance on the FIRST and NVSS radio maps. These included I, II, I/II, DD, X-shape, Z-shape, WAT, NAT, and HT–types. Apart from these formal radio AGN morphological classes, this catalog also included radio sources with one-sided FR morphologies, which were dubbed as O I and O II, respectively. The catalog provides halo-type morphology and star-forming region (SFR) morphology where the radio emission arises from star-forming regions of the galaxy. The catalog provides morphologies that are not clear (NC), blended (B; radio emission originated from more than one source), not detected (ND), where the radio cores were found to be associated with another nearby optical galaxy (see Figure 1 of D. Kozieł-Wierzbowska et al. 2020). D. Kozieł-Wierzbowska et al. also assigned 'possible' classification with the prefix p to the classes mentioned above whenever the 'secure' classification was unfeasible. The total number of sources (including possible classifications) was 4785 C, 24,252 E, 416 I, 871 II, 216 I/II, 14 X, 25 Z, 20 DD, 309 WAT, 125 NAT, 28 HT, 224 O I, 228 O II, 423 SFR, 18 NC, 414 B, and 39 ND types. Throughout this paper, we have adopted the same acronyms to describe radio morphologies of the radio AGNs as those given in Table 2 of D. Kozieł-Wierzbowska et al. (2020).

We extracted a total of 2,477 multiple-component radio AGNs belonging to the I, II, I/II, DD, X, Z, WAT, NAT, HT, O I, and O II classes from the ROGUE I catalog (D. Koziel-Wierzbowska et al. 2020). We note that a total 454 O I or O II radio sources identified in the ROGUE I catalog fall in the category of multiple-component radio AGNs; however, we do not consider them in the present analysis. This is because the existence of one-sided AGN classes remains debatable, mainly because two-sided jets are formed via the Blandford-Znajek and Blandford-Payne mechanisms (R. D. Blandford & R. L. Znajek 1977; R. D. Blandford & D. G. Payne 1982). However, there are a few possibilities for how one-sided radio AGNs can appear in the radio maps: 1) mis-identification of single-component radio AGN as O I or O II class because the lobe emission belongs to another back ground galaxy not detected in DSS maps, (2) extreme flux asymmetry of lobe and counter-lobe in the double-lobed radio AGN, (3) the mis-identification of double-lobed radio AGN as single-lobed because radio emission from one lobe overlaps on another optical galaxy present in the field, and (4) belong to HT-class but could not be identified as such because of lack of clear tail-like morphology due to limited sensitivity of to the extended emission of the FIRST survey. We can rule out the possibility that the lobe emission belongs to another optical galaxy in the field by examining optical images from the SDSS survey that have a deeper magnitude limit of $g' \sim 22.0$ (C. Stoughton et al.

[9] http://rogue.oa.uj.edu.pl/index.php?page=rogueI

**Table 1.** List of acronyms used and their meaning for size measurements, as well as the typical uncertainty in measured total angular size in arcsec (see Section 3.1 for details)

| Keyword | Meaning | Uncertainty |
|---|---|---|
| F | positions of the lobe and counter-lobe components detected in the FIRST catalog | 1.4 |
| N | positions of the lobe and counter-lobe components detected in the NVSS catalog | 9.9 |
| FN | positions of lobe component detected in the FIRST catalog and the counter-lobe component detected in the NVSS catalog | 7.1 |
| FMOS | position of the lobe component detected in the FIRST catalog, while the position of the counter-lobe measured *manually* from the FIRST map using DS9) | 2.9 |
| NMOS | position of the lobe component detected in the NVSS catalog while the position of the counter-lobe measured *manually* from the NVSS map using DS9) | 23.6 |
| FNMOS | position of the lobe component detected in the FIRST catalog while the position of the counter-lobe measured *manually* from the NVSS map using DS9 | 22.5 |
| NFMOS | position of the lobe component detected in the NVSS catalog while the position of the counter-lobe component measured *manually* from the FIRST map using DS9 | 7.5 |
| FMTS | positions of the lobe and the counter-lobe components measured *manually* from the FIRST map using DS9 | 3.8 |
| NMTS | positions of the lobe and the counter-lobe components measured *manually* from the NVSS map using DS9 | 31.8 |
| FDB | two radio components detected on FIRST map (see Appendix A for more information) | 1.4 |

2002), than the R-band DSS images of 20.8[10] used in the ROGUE I catalog. Since the double-lobed radio AGNs exhibit up to one order of magnitude flux ratios (Y. A. Gordon et al. 2023), the weaker sources with total luminosity $\leq$10 mJy may appear as single-lobe sources for modest to extreme flux asymmetry ratios for our sample. However, since the ROGUE I catalog does not provide the lobe and counter-lobe fluxes of radio AGNs, we cannot assess the likelihood that a one-sided AGN will appear for our sample. Moreover, the possibility that a double-lobed AGN is classified as single-lobe because the emission from one lobe falls on another optical galaxy can be excluded by inspection of spectral index maps made between lower frequency, preferably, 150 MHz and 1.4 GHz, along with the radio morphology. In this case, the spectral index $\alpha$ ($S \propto \nu^{-\alpha}$; $S$ and $\nu$ are flux density and frequency, respectively) will be steep ($\alpha \sim$1) for the lobes and it will be flat ($\alpha \lesssim$0.5) for the radio core (K. I. Kellermann & F. N. Owen 1988). Similarly, the possibility that one-sided sources are HT-types can be inferred by examining radio morphology with high-sensitivity, lower-frequency maps, where it should be more pronounced due to emission from older plasma. Ascertaining the radio AGN nature of the one-sided sources in the ROGUE I catalog is currently beyond the scope of this paper, in particular, because the Low Frequency Array (LOFAR)-Two metre Sky Survey (LoTSS) catalog for the entire northern sky is unavailable (the current data release (DR) 2 covers 27% of the LOFAR sky; T. W. Shimwell et al. 2022). In radio maps, we found that 23 of the 2477 multiple-component extended sources (ROGUE IDs: 609, 1027, 6161, 10537, 11612, 11614, 12220, 13185, 13261, 13287, 13928, 14350, 14389, 14691, 24529, 24840, 25264, 25499, 26526, 27564, 27608, 31253, and 32256) are described with single radio components and hence belong to C or E morphologies. This erroneous classification was an oversight, and we removed it from the current sample. This gives us a total of 2,002 multiple-component radio AGNs belonging to I, II, I/II, DD, X, Z, WAT, NAT, and HT-classes.

The availability of optical spectra and multi-filter ($u'$, $g'$, $r'$, $i'$, and $z'$) observations from the SDSS DR 7 for the ROGUE I galaxies allows us to compute specific physical properties such as the redshift, absolute $r-$band optical magnitude of the host galaxy, $M_r$, mass of black hole $M_{BH}$, total stellar mass ($M_\star$). The last two quantities are derived by fitting the inverse spectral synthesis code STARLIGHT to the optical spectrum (R. Cid Fernandes et al. 2005). At present, this is the most extensive handmade catalog of its kind and it provides an extremely rich data set to study the evolution of radio AGNs and star-forming galaxies with respect to (1) optical and radio morphology type (2) optical luminosity from the SDSS database, (3) radio luminosity of the core and total emission, and (4) black hole mass and stellar mass, and (5) inter-comparison of physical properties within the radio AGN morphological classes.

[10] https://www.cadc-ccda.hia-iha.nrc-cnrc.gc.ca/en/dss/

**Table 2.** Basic parameters of the first 20 multiple-component radio sources presented in the ROGUE I catalog

| ID | RA | Dec | z | Opt | Radio | $S_{core}$ | $eS_{core}$ | $S_{total}$ | $eS_{total}$ | Scale | RA(lobe) | Dec(lobe) | RA(c.lobe) | Dec(c.lobe) | Method | $LS_{total}$ | $\log_{10}(M_\star)$ | vd |
|---|---|---|---|---|---|---|---|---|---|---|---|---|---|---|---|---|---|---|
| | (deg) | (deg) | | | | (mJy) | (mJy) | (mJy) | (mJy) | (kpc/″) | (deg) | (deg) | (deg) | (deg) | | (kpc) | ($M_\odot$) | (km s$^{-1}$) |
| (1) | (2) | (3) | (4) | (5) | (6) | (7) | (8) | (9) | (10) | (11) | (12) | (13) | (14) | (15) | (16) | (17) | (18) | (19) |
| 13 | 146.78151 | 0.73795 | 0.2619 | E | WAT | 9.37 | 0.15 | 49.30 | 1.90 | 4.0824 | 146.7769 | 0.7419 | 146.7798 | 0.7314 | F | 188.7 | 12.05 | 290.0 |
| 18 | 149.16988 | -0.02335 | 0.1393 | E | II | 2.15 | 0.14 | 199.80 | 5.23 | 2.4759 | 149.1613 | -0.0125 | 149.1795 | -0.0348 | F | 256.6 | 11.62 | 256.0 |
| 24 | 147.70857 | -0.88773 | 0.2715 | E | II | 18.95 | 0.13 | 59.80 | 2.20 | 4.1900 | 147.7136 | -0.8876 | 147.7052 | -0.8879 | F | 126.8 | 11.94 | 304.4 |
| 25 | 147.42824 | -0.84007 | 0.0810 | E | I/II | 11.69 | 0.15 | 177.70 | 6.40 | 1.5383 | 147.4234 | -0.8379 | 147.4319 | -0.8464 | F | 69.9 | 11.07 | 182.5 |
| 67 | 152.45076 | 0.48414 | 0.1859 | E | II | 5.40 | 0.16 | 35.80 | 1.50 | 3.1404 | 152.4489 | 0.4888 | 152.4531 | 0.4813 | F | 98.3 | 11.33 | 222.9 |
| 90 | 154.90808 | 0.33215 | 0.0956 | E | pI/II | 10.48 | 0.14 | 48.40 | 1.87 | 1.7848 | 154.9152 | 0.3247 | 154.8885 | 0.3453 | FN | 217.8 | 11.61 | 266.3 |
| 128 | 159.02403 | 0.10189 | 0.0968 | E | II | 109.92 | 0.14 | 544.60 | 14.08 | 1.8054 | 159.0425 | 0.1150 | 159.0113 | 0.0895 | F | 262.7 | 11.82 | 309.0 |
| 152 | 161.40544 | 0.02123 | 0.4417 | E | pI | 5.25 | 0.15 | 13.70 | 1.00 | 5.7604 | 161.4016 | 0.0271 | 161.4077 | 0.0164 | F | 256.0 | 11.96 | 296.8 |
| 157 | 160.38312 | -0.45925 | 0.1374 | E | I/II | 6.38 | 0.14 | 12.70 | 1.10 | 2.4476 | 160.3843 | -0.4576 | 160.3817 | -0.4612 | FMOS | 39.1 | 11.55 | 263.9 |
| 164 | 162.30870 | 0.99592 | 0.1065 | E | I/II | 3.13 | 0.14 | 99.80 | 2.69 | 1.9640 | 162.3195 | 1.0006 | 162.3001 | 0.9961 | F | 144.0 | 11.55 | 257.6 |
| 178 | 162.49522 | 0.32230 | 0.0390 | E | II | 6.00 | 0.18 | 61.80 | 2.40 | 0.7783 | 162.4922 | 0.3255 | 162.4986 | 0.3186 | F | 26.4 | 11.42 | 262.8 |
| 190 | 165.06720 | -0.86816 | 0.1580 | E | pII | 0.93 | 0.14 | 92.50 | 3.06 | 2.7520 | 165.0697 | -0.8547 | 165.0572 | -0.8810 | FN | 296.9 | 11.60 | 208.6 |
| 263 | 171.84988 | 0.13287 | 0.1338 | E | WAT | 5.26 | 0.15 | 12.40 | 0.60 | 2.3935 | 171.8506 | 0.1361 | 171.8465 | 0.1331 | F | 57.7 | 11.28 | 261.4 |
| 274 | 172.64439 | 0.88072 | 0.1350 | E | NAT | 7.84 | 0.15 | 24.09 | 0.26 | 2.4114 | 172.6366 | 0.8824 | 172.6371 | 0.8756 | F | 146.5 | 11.41 | 262.1 |
| 311 | 177.93062 | 1.07891 | 0.2665 | E | pII | 28.09 | 0.16 | 41.02 | 0.23 | 4.1346 | 177.9307 | 1.0805 | 177.9303 | 1.0788 | FDB | 28.7 | 11.68 | 268.6 |
| 313 | 178.63919 | 0.96693 | 0.1574 | E | pII | 1.58 | 0.15 | 11.00 | 1.30 | 2.7433 | 178.6308 | 0.9659 | 178.6440 | 0.9671 | FMOS | 131.0 | 11.37 | 183.3 |
| 322 | 179.92345 | -0.53166 | 0.1783 | E | II | 2.80 | 0.14 | 57.30 | 2.28 | 3.0365 | 179.9149 | -0.5255 | 179.9373 | -0.5396 | N | 289.7 | 11.81 | 260.8 |
| 333 | 179.58873 | 0.61837 | 0.2720 | E | II | 6.39 | 0.14 | 22.30 | 1.30 | 4.1954 | 179.5906 | 0.6259 | 179.5856 | 0.6131 | F | 209.8 | 11.74 | 274.3 |
| 358 | 182.22730 | 1.01775 | 0.3485 | E | pII | 1.80 | 0.15 | 6.50 | 0.50 | 4.9752 | 182.2266 | 1.0182 | 182.2284 | 1.0169 | FDB | 39.8 | 11.59 | 236.1 |
| 362 | 184.00227 | -0.67232 | 0.3442 | E | II | 4.48 | 0.16 | 13.40 | 1.20 | 4.9346 | 184.0050 | -0.6719 | 183.9987 | -0.6725 | FMOS | 112.6 | 11.88 | 282.7 |

Note—The table is available in its entirety in the machine-readable form. (1) ROGUE identification number, (2–4) RA, dec. and redshift of the optical host galaxy from the SDSS database, (5) optical morphological classification of the optical host from the SDSS image given in the ROGUE I catalog, (6) final radio morphological classification of the associated radio source given in the ROGUE I catalog, (7–8) 1.4 GHz core flux density and the corresponding error of the radio source from the FIRST database, (9–10) 1.4 GHz total flux density and the corresponding error of the radio source, (11) angular to linear scale coversion factor, (12–13) RA and dec. of the farthest lobe component, (14–15) RA and dec. of the farthest counter-lobe component, (16) method of lobe position estimation: F-RA and dec. of the radio component from the FIRST database (columns 12, 13, 14, and 15), N-RA and dec. of the radio component from the NVSS database (columns 12, 13, 14, and 15), FN-RA and dec. from FIRST database for columns 12 and 13 and from NVSS database for columns 14 and 15, FDB-RA and dec. of the radio components from the FIRST database (columns 12, 13, 14, and 15), FMOS-RA and dec. from FIRST database for columns 12 and 13 and manually measured from the FIRST map for columns 14 and 15; NMOS-RA and dec. from NVSS database for columns 12 and 13 and manually measured from the NVSS map for columns 14 and 15; FMTS-RA and dec. manually measured from the FIRST map (columns 12, 13, 14, and 15), and NMTS-RA and dec. manually measured from the NVSS map (columns 12, 13, 14, and 15) (see Section 3.1 for details), (17) total linear size, (18) galaxy mass from the STARLIGHT database, and (19) stellar velocity dispersion from the STARLIGHT database.

## 3. ESTIMATION OF PHYSICAL PROPERTIES

### 3.1. *Projected Linear Sizes*

The angular size of the radio AGN is computed in a standard fashion: it is a sum of straight-line distances from the position of the optical host galaxy to the positions of farthest radio components belonging to the AGN, i.e., the hot spot/lobe and the counter-hot spot/lobe, respectively. The angular distance from the optical host to the hot spot/lobe, $D_{\rm lobe}$, is computed as:

$$D_{\rm lobe} = \sqrt{((\alpha_{\rm opt} - \alpha_{\rm rad}) \times \cos\delta_{\rm opt})^2 + (\delta_{\rm opt} - \delta_{\rm rad})^2} \quad (1)$$

where $\alpha_{\rm opt}$, $\delta_{\rm opt}$ are the right ascension and declination of the optical positions of the optical host galaxy from the SDSS DR 7, respectively, and $\alpha_{\rm rad}$, $\delta_{\rm rad}$ correspond to the right ascension and declination of the radio positions of the lobe from either the FIRST or the NVSS surveys, respectively. Similarly, the angular distance from the optical host to the counter-lobe is computed. The distances to the lobe and the counter-lobe are multiplied by the angular-to-linear scale conversion factor at the source distance, and summed to give the projected total linear size. For DDs and Z-shaped sources, the distance to the farthest radio component is taken as a measure of source size, whereas for X-shaped sources, the distance to the radio component with the same pair of jets/outflows with the largest separation is taken as a measure of size.

As the ROGUE I catalog provides the final radio morphology by combining information from the FIRST and NVSS maps, which have different resolutions and sensitivity limits for extended and point-like emission, measuring the positions of the farthest radio components for the size measurement proved to be non-trivial. Below, we briefly describe our methodology for locating the farthest radio lobe components for the sources, while Figure 1 illustrates examples, and Table 1 gives the meaning of the keywords assigned.

For the size measurements, we preferred the radio component positions from the FIRST survey over those from the NVSS survey. If a given radio component was detected in both the FIRST and NVSS catalogs, we used the position yielding the largest distance from the optical galaxy. If the radio component was not detected on either map, we manually measured it using DS9 (i.e., the emission above the detection threshold of the FIRST and NVSS catalogs), or it presents an emission too small to be identified as an individual component in the radio maps. In about 78% of cases, we use positions of radio components (peak of the fitted 2-dimensional Gaussian) from the FIRST or NVSS databases to measure the sizes, while for ∼22% cases, the sizes are measured manually on the maps.

The FIRST and the NVSS surveys give astrometric accuracies in the range of ∼0.5–1″ (R. H. Becker et al. 1995) and ∼1–7″(J. J. Condon et al. 1998), respectively, with brighter source positions determined with better accuracy. Since the ROGUE I catalog was compiled without flux-density limits for radio-component detections, we assign the largest astrometric uncertainties to their positions, i.e., 1″ for the FIRST and 7″ for the NVSS surveys. For manually measured cases, positions can be measured with high accuracy if the radio emission increases outwards (e.g., FR type-II), but would have higher uncertainty if the intensity is monotonically decreasing (e.g., FR type-I). Conservatively, for the manually measured cases, we assign an uncertainty in angular sizes, assuming the worst case, meaning that sizes are overestimated by half the beam size. The astrometric accuracy of the optical positions of the host galaxy from the SDSS survey is ∼100 mas (J. R. Pier et al. 2003). The uncertainties in the optical position and the radio position of the lobe components are added in quadrature to obtain the uncertainty in the total angular size. The last column of Table 1 gives these uncertainties, which can be converted to uncertainties in the total linear sizes using the angular to linear conversion factors at the source distance.

We flag 81 sources as 'FDB' in column 16 of Table 2. These sources show only two radio components in the FIRST map, separated by nearly equal distances from the optical galaxy. Their radio morphology on the FIRST map appeared to be that of classical radio sources, i.e., sources with a radio core and two radio lobes. Therefore, the majority of them were classified as 'possible' candidates of different morphologies. Specifically, 13 type-I, 52 type-II, six type-I/II, and one WAT type. 9 out of 81 sources were assigned secure classification (2 type-I, 5 type-II, and 2 WAT). As these sources do not present a clear radio core component in the FIRST maps, in Appendix A, we further examined their high-resolution (2.5″) radio maps made at 3.0 GHz using the data from the VLA Sky Survey (VLASS; M. Lacy et al. 2020). About 25% of the sources detected in the VLASS maps show a distinct radio core, confirming that the FIRST-detected components are associated with the lobes, while the rest do not reveal a clear core. To ensure uniformity of the analysis procedure, we keep their sizes as measured from the positions of the lobe components on the FIRST maps.

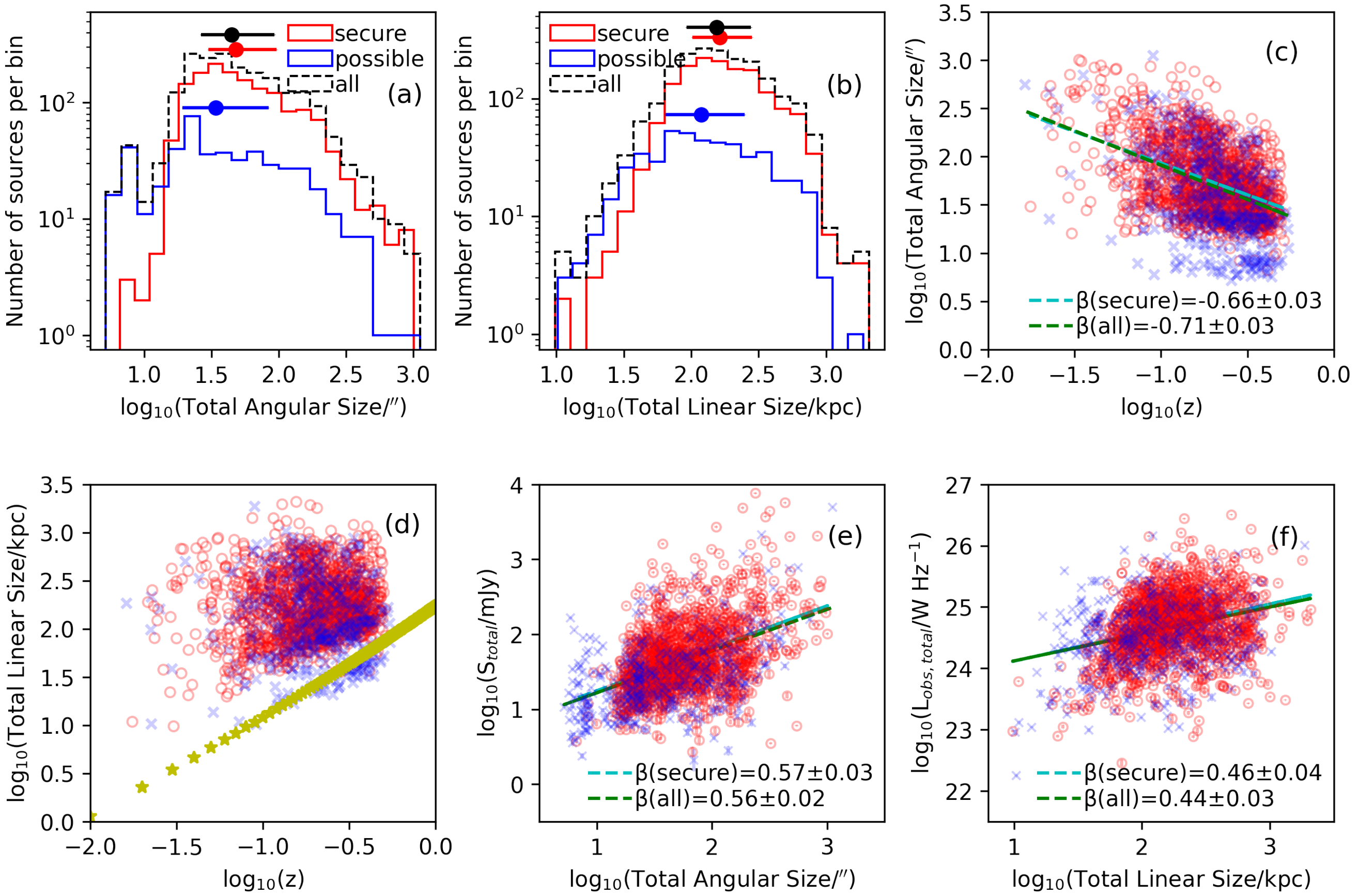

**Figure 2.** Properties of radio AGN population studied here: (a) histogram of total angular sizes, (b) histogram of total linear sizes, (c) distribution of total angular size with $z$, (d) distribution of total linear size with $z$, (e) distribution of 1.4 GHz total flux density with total angular size, and (f) distribution of 1.4 GHz total luminosity with total linear size. The red and blue colors show the sources with 'secure' (1527 sources) and 'possible' (475 sources) identifications, respectively. The solid lines in red and blue and the dashed line in black give distributions considering 'secure', 'possible', and 'all' (secure+possible) identifications. The horizontal segments on top of the histograms indicate the 16th and 84th percentiles, with a filled circle marking the 50th percentile (panels a and b). $\beta$(secure), shown by a cyan dashed line, is the index of the power-law fitted between the quantities using the sample of sources with secure morphological identifications only, while $\beta$(all), shown by a green dashed line, is the index using the full sample of sources, i.e., secure+possible morphological identifications, while the error on $\beta$ is the standard deviation. Yellow stars in panel e mark the linear sizes corresponding to the beam size of the FIRST survey (5.4″).

### 3.2. *Measurement of luminosities, absolute optical magnitude, and black hole masses*

Besides giving the optical and radio morphological classifications, the ROGUE I catalog provides the 1.4 GHz core and total flux densities ($S_{\rm core}$ and $S_{\rm total}$ hereafter) of the radio sources. The luminosity in the observer's frame, $L_{\rm obs}$, is related to the flux density, $S$, as:

$$L_{\rm obs} = 4\pi {d_{\rm L}}^2\, S\, [{\rm W\,Hz^{-1}}] \quad (2)$$

where $d_{\rm L}$ is the luminosity distance computed using Eq. 2 of D. Kozieł-Wierzbowska et al. (2020). We also compute the extinction corrected $r-$band absolute magnitude, $M_r$, from the apparent $r-$band magnitude, $m_r$, given in the ROGUE I catalog:

$$M_r = m_r - A_r + 5 - 5\,\log_{10}(d_{\rm L}), \quad (3)$$

where $A_r$ is the galactic extinction coefficient provided by D. J. Schlegel et al. (1998) and $d_{\rm L}$ (provided in the ROGUE I) is measured in units of pc.

The SMBH mass, $M_{\rm BH}$, is computed using the stellar velocity dispersion, $\sigma_*$ which is determined from the STARLIGHT program (R. Cid Fernandes et al. 2005) and following the relation by S. Tremaine et al. (2002):

$$\log_{10}(M_{\rm BH}/M_\odot) = 8.13 + 4.02\,\log_{10}(\sigma_*/200\,{\rm km\,s^{-1}}) \quad (4)$$

Additionally, we obtained stellar mass, $M_\star$ from the STARLIGHT database (R. Cid Fernandes et al. 2005).

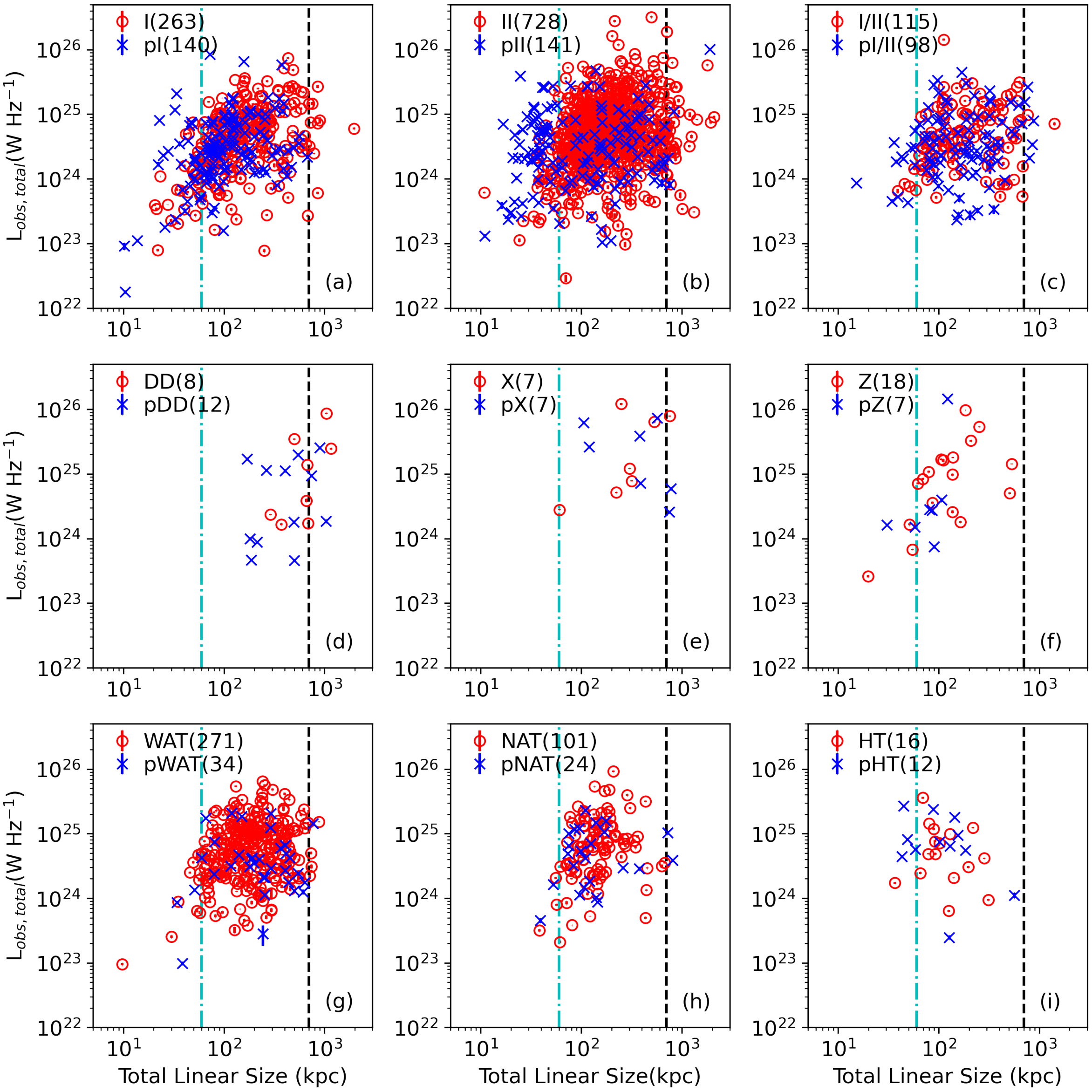

**Figure 3.** Observed 1.4 GHz total luminosity plotted against the projected total linear size for I, II, I/II, DD, X, Z, WAT, NAT, HT-classes of radio AGNs. The red and blue points show the sources with 'secure' and 'possible' identifications, respectively, and the possible identifications of the morphological class are indicated by adding a prefix p. In parentheses, we note the number of sources belonging to each morphological class. Vertical lines at 60 kpc (dot-dashed cyan) and 700 kpc (dashed black) are drawn in each panel for visual aid, marking the compact resolved AGNs and the giant radio AGNs, respectively. We note that the total linear size for the HT class (panel i) corresponds to the size from the optical core to the lobe on one side (see Table 2).

## 4. RESULTS AND DISCUSSION

### 4.1. *Angular/projected linear sizes and their relations with z, total flux density and radio power*

Figure 2 gives the histograms of total angular and projected linear sizes (panels a and b), the distributions of total angular size/projected linear size with $z$ (panels c and d), distribution of $S_{\rm total}$ with total angular size (panel e), and $L_{\rm total}$ with total linear size (panel f) for the radio AGNs studied here. Figure 3 gives the distributions of projected linear sizes and total radio luminosity for each morphological class.

#### 4.1.1. *Distribution of total angular/projected linear sizes*

Figures 2a and 2b show histograms of the total angular and projected linear sizes of the radio AGNs. We find that extended radio sources have angular sizes in the range of $\sim5''$ to $\sim1100''$, with the majority showing angular sizes of $\leq100''$ (median$\sim45''$ for all sources). Similarly, the projected linear sizes range from $\sim10$ kpc to 2,200 kpc, with the majority of sources having sizes <200 kpc (median∼154 kpc for all sources). Furthermore, the angular/linear size distributions are similar when comparing histograms for secure and possible radio morphological identifications presented in the ROGUE I catalog (panels a and b of Figure 2), indicating a lack of bias when measuring the sizes. We note that the range of linear sizes exhibited by extended radio AGNs at 1.4 GHz is similar to that obtained at 150 MHz in the LoTSS DR 2 data (M. J. Hardcastle et al. 2025).

We find that 211 radio AGNs have a total projected linear size≤60 kpc. These are equivalent to the compact radio source catalog –*COMP2CAT*– presented by A. Jimenez-Gallardo et al. (2019). 96 of these sources correspond to a secure radio morphological identification, while 115 sources belong to possible identifications of different morphological classes. The very small sizes of these sources indicate that the lobes are still expanding within the optical host galaxies. They present a sample of AGN that allows study of the feedback processes that affect host galaxy evolution by impacting their immediate environments on scales up to ∼100 kpc. The energy carried by jets heats the gas in the interstellar medium (ISM), which stops them from forming stars, which in turn affects the growth of the central AGN or the positive feedback which induces starburst in galaxies (e.g., D. J. Croton et al. 2006; D. Wylezalek & N. L. Zakamska 2016; M. Valentini et al. 2020). However, we do not disregard the possibility that in these sources the jets could be confined, as the growth can be frustrated, as seen in much smaller size ≤15 kpc compact symmetric objects (C. P. O'Dea 1998) or the jets can be short-lived (M. Kunert-Bajraszewska et al. 2010). We will discuss them in detail in our forthcoming study.

We also find a total of 71 'giant' radio galaxies with total projected linear size >700 kpc (A. Kuźmicz et al. 2018; P. Dabhade et al. 2020a). These sources are important for understanding why only a small fraction of radio AGNs tend to grow to very large sizes (S. Sankhyayan & P. Dabhade 2024; M. Simonte et al. 2024) and AGN feedback on intracluster scales (M. J. Hardcastle et al. 2007; P. N. Best et al. 2007; B. R. McNamara & P. E. J. Nulsen 2012).

Moreover, Figure 3 shows that the detection of small-size (≤60 kpc) and giant (≥700 kpc) radio AGNs is not restricted to any particular morphological AGN class.

#### 4.1.2. *Distribution of angular/linear sizes as a function of z*

Figures 2c and 2d show the distribution of total angular and projected linear sizes as a function of $z$. We observe an inverse relation between the angular size and $z$, which is expected because of the inverse relation between the angles and the distance in a flat expanding Universe. We perform the Spearman rank correlation test to measure the strength ($p$-value) and direction (positive or negative) given by the $\rho$ value of the test. We obtain $\rho = -0.41$ and $p < 10^{-5}$ for a sample of sources with secure identifications only, and it changes to $\rho = -0.42$ and $p < 10^{-5}$ for a full sample of sources with secure+possible identifications (Figure 2c). The $p-$values indicate that the correlation is statistically significant at confidence levels ≥99.7%, and is moderately strong ($\rho$ ∼-0.4). We fit a power-law between total angular size and redshift using linear regression and find that the power-law index, $\beta \sim -0.7$, within 1$\sigma$ uncertainties (Figures 2c).

The Spearman rank correlation test between the linear size and $z$ indicates $\rho$ and $p-$values of 0.1 and $<10^{-4}$ respectively, for a sample of sources with secure identifications only. The $\rho$ and $p-$values change to 0.05 and 0.03, respectively, for a full sample of sources with secure+possible identifications (Figures 2d). The small values of $\rho$ ($< 0.1$) imply a lack of correlation of linear sizes with $z$.

#### 4.1.3. *Distribution of 1.4 GHz total radio flux density as a function of total angular size*

Figure 2e gives the distribution of 1.4 GHz total flux density with angular size. The Spearman rank correlation test gives $\rho = -0.47$ and $p < 10^{-5}$, respectively, for a sample of sources with secure identification only, which changes to $\rho = -0.44$ and $p < 10^{-5}$, respectively, for a full sample of sources with secure+possible identifications. The $p-$values indicate that the correlation is statistically significant at confidence levels ≥99.7% and is moderately strong ($\rho$ ∼0.4). We fitted a power law using linear regression and found that the flux density varies with angular size raised to the power ∼0.6.

In a flat Universe, since the flux density varies as the inverse square of distance, and the angular size is inversely proportional to the distance, one expects flux density to vary as the square of the angular size. Our finding indicates that flux density varies with angular size, with an index that is approximately 3.5 times lower than expected. Since the largest angular size sources are located at the lowest redshifts (Figure 2c), our result indicates a lack of bright sources in lower redshifts.

#### 4.1.4. *Distribution of 1.4 GHz total radio power as a function of projected total linear size*

Figure 2f provides the distributions of 1.4 GHz total radio luminosity with total linear size. The Spearman rank correlation test give $\rho$ and $p-$values of 0.27 and $< 10^{-5}$, respectively, for a sample of sources with secure identifications only, which change to 0.26 and $< 10^{-5}$ for a full sample of sources with secure+possible identifications. The $p-$values indicate that the correlation is statistically significant at confidence levels greater than 99.7% and is moderately strong ($\rho$ ∼0.3). The result of the power-law fitting using linear regression indicates that the total radio power scales as ∼0.5 power of the projected linear size, slightly steeper than the reported index of 0.3 in the literature (see, K. M. Blundell et al. 1999).

| Source Type | redMaPPer | W12 | Total | ROGUE I | Fraction 1 | $n_{200c}$ (Mpc$^{-3}$) | Asy. | Fraction 2 |
|---|---|---|---|---|---|---|---|---|
| (1) | (2) | (3) | (4) | (5) | (6) | (7) | (8) | (9) |
| I(pI) | 5(2) | 73(39) | 78(41) | 263(140) | 29.6%(29.2%) | 3.4±0.8 (3.5±0.7) | 53(33) | 20.1%(23.5%) |
| II(pII) | 28(7) | 203(35) | 231(42) | 728(141) | 31.7%(29.8%) | 3.8±1.5 (3.8±0.9) | 53(41) | 7.4%(29.1%) |
| I/II(pI/II) | 3(3) | 29(25) | 32(28) | 115(98) | 27.8%(28.9%) | 3.9±0.9 (3.8±0.9) | 27(31) | 23.5%(31.6%) |
| DD(pDD) | 0(0) | 1(1) | 1(1) | 8(12) | 12.5%(8.33%) | 3.3±0.0 (3.7±0.0) | 1(2) | 12.5%(16.7%) |
| X(pX) | 0(0) | 1(3) | 1(3) | 7(7) | 14.3%(42.8%) | 3.0±0.0 (3.2±0.3) | 0(1) | 0%(14.2%) |
| Z(pZ) | 2(0) | 6(4) | 8(4) | 18(7) | 44.4%(57.1%) | 3.2±0.8 (3.4±1.1) | 2(1) | 11.1%(14.2%) |
| WAT(pWAT) | 17(1) | 100(11) | 117(12) | 271(34) | 43.2%(35.3%) | 3.7±0.9 (3.3±0.6) | 35(11) | 12.9%(32.3%) |
| NAT(pNAT) | 6(1) | 53(11) | 59(12) | 101(24) | 58.4%(50%) | 3.4±0.6 (3.6±0.4) | 16(6) | 15.8%(25.0%) |
| HT(pHT) | 2(0) | 7(4) | 9(4) | 16(12) | 56.3%(33.3%) | 3.8±0.6 (3.1±0.3) | - | - |

**Table 3.** Number of radio AGNs with secure identification with different radio morphologies found in the cluster environments. In parentheses, we give the same, but for possible identification of a given radio morphological class. Columns are: (1) source type, (2) number of sources found in the redMaPPer unique cluster catalog, (3) number of sources found in the Z. L. Wen et al. (2012) catalog, (4) total number of sources found in the cluster environment (column 2+ column 3), (5) total number of sources in the ROGUE I catalog, (6) fraction of sources found in the cluster environment (column 4/column 5), (7) mean number density of member galaxies with $1\sigma$ standard deviation within $R_{200c}$ of the associated clusters found in the Z. L. Wen et al. (2012) catalog, (8) number of sources with arm-length ratio $\geq$2 in our sample, except for the HT sources, (9) fraction of sources with with arm-length ratio $\geq$2 in our sample (column 8/column 5).

Next, we fitted 2D linear regression assuming that linear size varies as a function of redshift and total power. We find that the linear size scales as $z^{-0.14}$ and varies as the total radio power raised to 0.23 for a sample of sources with secure identifications only, and scales as $z^{-0.26}$ and is proportional to the total radio power raised to 0.28, for a full sample of sources with secure+possible identifications. The 2D power-law fit of linear size (redshift, total power) indicates that small size sources are also radio-weak for a constant redshift.

#### 4.1.5. *Implications and Limitations*

Sections 4.1.2, 4.1.3, and 4.1.4 quantify the relations between angular/linear sizes with $z$, flux density with angular size, and radio power with linear size for the extended radio AGNs. In the relativistic cosmological models where the geometry of space-time is described by the Friedmann-Robertson-Walker metric, the angular size of the object decreases inversely with redshift up to a maximum redshift of 1.6 (as in Euclidean space), beyond which it starts to grow again. Since the maximum $z$ for our sample is 0.54, we expect a strict inverse relationship between the angular size and $z$ for a flat Universe; however, the index of the fitted power-law is smaller than -1 (∼-0.7; Figure 2c ). Similarly, the index of power-law fit between flux density and angular size is smaller than 2 (∼0.6; Figure 2e). Linear size does not seem to vary with redshift (Figure 2d). Since any sample can be affected by Malmquist bias (K. G. Malmquist 1922) due to the sensitivity limit of an imaging instrument, fainter sources are less likely to be detected at higher redshifts. We examined whether any of the observed trends using the full sample of radio AGNs are affected by it, and we present the results in Appendix B. We observe similar trends (within errors) in our volume-limited sample as in the full sample: between angular/linear size and z, and between flux density and z. The index of the power-law fit between total radio power and total linear size is slightly flatter, ∼0.2, than in the full sample, and is consistent within errors with the value reported in the literature. So, although the Malmquist bias is present in our sample, the observed trends are not dominated by it, partly because of the limited $z$ range of 0.01–0.54, with the exception of the total radio power vs. linear size trend.

### 4.2. *Large scale environments*

Radio AGNs of different morphological types have been used to probe the large-scale environments in which they reside, such as galaxy groups or clusters. For example, it is hypothesized that the bent-angle sources (WAT, NAT, and HT types) are formed from the straight-line sources (I, II, and I/II types) due to the increased ram pressure experienced by the lobes as the galaxy moves through the intragroup/intracluster medium (T. W. Jones & F. N. Owen 1979; A. F. Garon et al. 2019; K. de Vos et al. 2021), or the interaction of lobes with gaseous halo/ISM of galaxies in the cluster medium (J. T. Stocke et al. 1985), or due to gas sloshing in their intracluster medium due to cluster mergers (I. Sakelliou & M. R. Merrifield 2000; R. Paterno-Mahler et al. 2013). Regardless of the physical mechanism that causes the radio lobes to bend, dense environments appear essential for their formation. And indeed, it has been shown in previous studies, that a significant fraction of these sources reside in galaxy groups/cluster environments (F. N. Owen & L. Rudnick 1976; E. M. Silverstein et al. 2018; B. Mingo et al. 2019; M. E. Mor-ris et al. 2022).

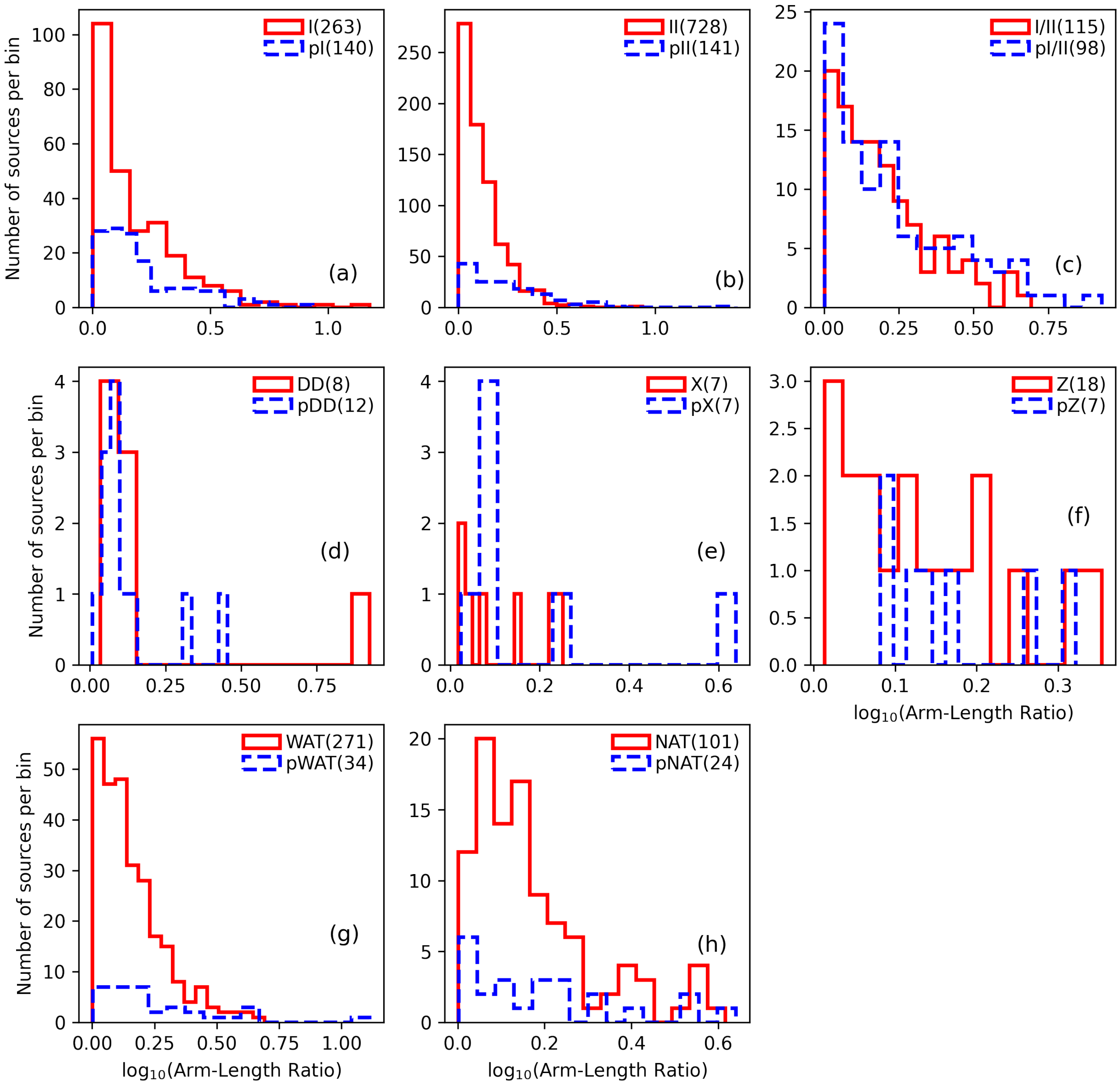

**Figure 4.** Histograms of arm-length ratios (i.e., longer lobe to the shorter counter-lobe), for the radio AGN classes. The red and blue lines give the distribution for the secure and possible identifications, respectively, and the possible identifications of the morphological class are indicated by adding a prefix p. The name of the radio morphological class and the number of sources included in the distribution are given in each panel. The distributions are sharply peaked at 1, indicating that the sizes of the lobe and the counter-lobe are generally similar across all morphological classes.

We checked what fraction of our sample lies in cluster/galaxy group environments by cross-matching with the two cluster catalogs made using the SDSS data, i.e., redMaPPer catalog by E. S. Rykoff et al. (2014) and Z. L. Wen et al. (2012) catalog (W12 hereafter). The redMaPPer catalog is based on SDSS DR8, covering an area of ∼10,000 deg$^2$ and contains ∼25,000 clusters in the $z$ range 0.08–0.55. The W12 catalog is based on SDSS DR3 and covers the area of ∼14,000 deg$^2$ and contains 132,684 clusters in the $z$ range 0.05–0.42. We note that despite covering roughly a similar area of the sky and using SDSS as a base catalog, the number of clusters identified in W12 catalog is a factor of 5 larger than in the redMaPPer catalog. This is because W12 catalog extends to lower richness in clusters. For cross-matching of radio sources with cluster catalogs, we used a radius of 1 Mpc from the cluster center with a $z$ error of 0.01, which corresponds to ∼3 times the typical velocity dispersion of a massive cluster. During the cross-matching process, a total of 303 radio AGN were associated with clusters in the redMaPPer catalog and 734 with clusters in the W12 catalog. Since both cluster catalogs cover similar areas on the sky, they have common clusters. Therefore, in order to remove common sources found in these cluster catalogs, we further matched between these two cross-matched subsets, based on the radio AGN right ascension and declination. This yielded 211 sources identified as cluster members in both catalogs. These common sources were removed from the redMaPPer catalog, leaving unique sources associated with it. The results are provided in Table 3.

We find that ∼34% of our sample of radio AGNs are associated with cluster environments (Column 6; Table 3). We note that the cluster association fractions of radio AGN population in the LoTSS DR 1 and DR 2 are ∼10% (23,344 sources; J. H. Croston et al. 2019) and ∼17% (26,577 sources; T. Pan et al. 2025), respectively, much lower than the 34% found here. We attribute this discrepancy to the fact that the ROGUE I catalog is based on visual inspection of the morphology, while the LoTSS DR1 and DR2 data used automated methods to assign morphology for the majority of their sources, which can result in lower identification of radio lobes belonging to the host galaxy, if the radio emission does not extend all the way from the radio core to the lobes; hence the lowering the number of sources with FR morphologies.

We find that cluster association fractions vary between ∼10% and ∼60% across the morphological classes (Column 6; Table 3), with the bent-angle sources showing the highest cluster association fractions (∼40%-60%). We note that B. Mingo et al. (2019) found ∼48% of WAT and ∼49% of NAT sources in the LoTSS DR 1 are associated with the cluster environments, similar to the fraction obtained here.

Next, we examine the statistical significance of bending in the opening angle of the radio lobes to the environmental metric, i.e., their cluster association fraction. For this, we combine the straight-line sources in one population (I, II, and I/II) and bent-angle sources into another population (WAT, NAT, and HT). A WAT/NAT source could appear as a straight-line FR source if the radio galaxy axis, defined as the line bisecting the lobe opening angle, falls along our line of sight. For all other orientations of the radio galaxy axis, the bent-angle source would appear as a bent-angle source only; the observed bending angle could be either increased or decreased by projection. Therefore, our approach allows us to minimize contamination between the straight-line and bent-angle classes' source counts due to projection effects. Moreover, since we expect sources to be isotropically distributed, a much smaller fraction of WAT/NAT sources will appear as a straight-line source due to projection effects. Therefore, our approach of combining straight-line sources into one class and bent-angle sources into another yields a proper estimate of the environmental metric (i.e., cluster association fraction) as a function of the bending angle. We find that the mean and standard deviations of the cluster association fractions are 29.7%±1.9%(29.3%±0.45%) and 52.6%±8.2%(39.5%±9.1%), for the secure(possible) identifications within the straight-line and bent-angle classes, respectively. The cluster association fraction differs at ≥95% confidence levels, suggesting that bent-angle sources do indeed preferentially reside in the cluster environment relative to straight-line sources, implying that their bent-morphology is likely due to the denser environment. We note that the redshift-dependent cluster detection rate, as a function of cluster mass $M_{200}$ (the mass within a radius at which the mean density of a cluster is 200 times the critical density of the Universe) and richness parameters, may influence our cluster association fraction. For the SDSS clusters, the detection rate varies between 100% for $M_{200}>2\times10^{14}$ (and richness >34) up to $z\sim0.5$ and 75% for clusters with $M_{200}>0.6\times10^{14}$ (and richness >12) up to $z\sim0.42$, respectively, and drops significantly for higher $z$ (Z. L. Wen et al. 2012). Therefore, the redshift-dependent cluster mass/richness completeness may influence the reported cluster association fractions, although this effect should be modest given the limited redshift range of our sample.

We calculated the galaxy number densities for clusters associated with our sample. Column 7 of Table 3 lists the mean and standard deviation of $n_{200c}$ clusters in W12 catalog. $n_{200c}$ is the galaxy number density of the cluster within $R_{200c}$ of a cluster, where $R_{200c}$ is the radius of the cluster enclosing a mean density 200 times the critical density of the Universe. We find that the mean galaxy number densities of clusters associated with radio AGNs in the I, II, I/II, WAT, NAT, and HT classes are comparable. We note that $R_{200c}$ and the number of member galaxy candidates within $R_{200c}$ ($N_{200c}$) values required for the calculation of the number density $n_{200c}$, which are not available for the redMaPPer unique cluster catalog. However, only ∼10 % of the clusters associated with our sources (Column 2; Table 3) belong to redMaPPer unique cluster catalog. Therefore, excluding them from calculating the mean galaxy number densities of clusters does not affect our conclusion.

Our result that FR type-I and type-II radio galaxies are inhabited by clusters that have similar galaxy number densities is consistent with the results obtained by F. Massaro et al. (2020), who found that their FR type-I and type-II radio sources reside in clusters with similar richness parameters. Next, we find that the mean galaxy number density of clusters inhabiting the bent-angle sources is comparable. Our results corroborate those of E. Vardoulaki et al. (2021), who analyzed 108 bent-angle sources in the VLA-COSMOS field and found no correlation between bending angle and large-scale environment metrics, such as comoving volume densities or galaxy cluster masses. Furthermore, the mean galaxy number density of bent-angle sources is similar to that of straight-line FR sources, though the former are expected to reside in dynamically active or merging clusters (F. N. Owen & L. Rudnick 1976; E. L. Blanton et al. 2001). Our finding hints that the cluster richness/number density is not the sole reason for their bent-angle structures, rather local conditions related to the intra-cluster environment, such as subcluster motion and gas dynamics inside a cluster, play a crucial role, as suggested by M. Bliton et al. (1998).

### 4.3. *Asymmetry in arm lengths*

Since the bipolar jets ejected from the black hole/accretion disk systems are symmetric and they propagate in the IGM with the same thrust, sources should result in symmetric lobe structures, unless external factors influence them. The physical reasons why asymmetries in the arm-lengths can arise: (1) the properties of the plasma during the time of ejection from the SMBH-accretion disk system are different (confinement, velocity, etc.), (2) due to the projection effects, the differences in the light travel times (for the lobe and the counter-lobe) can make one side appear longer than the other, and (3) due to differences in the environments in which the lobe and counter-lobe propagate, i.e., one lobe propagating in a much denser environment than the other. The third scenario has been confirmed observationally, where arm-length asymmetries are anti-correlated with galaxy clustering properties (Radio Galaxy Zoo; P. E. Rodman et al. 2019) and have been revealed in 3D hydrodynamical simulations of jets propagating in cluster environments (P. M. Yates-Jones et al. 2021).

Figure 4 gives the distribution of the arm-length ratio defined as the ratio of the longer arm to the shorter arm lengths, for the two-sided radio AGN classes. In general, the distributions are sharply peaked at 1, indicating that the arm lengths are similar for each radio AGN morphological class. The vast majority of our sources show arm-length asymmetries of less than one order of magnitude, with the exception of only 3 sources showing larger values. Arm-length asymmetries in double-lobe radio AGNs up to one order of magnitude have been detected previously, for example, using the FIRST survey data (sample size=84; L. Lara et al. 2004), VLASS survey data (sample size=∼17,000; Y. A. Gordon et al. 2023), LOFAR data (sample size=215; M. Mahato et al. 2025), and the TIFR-GMRT Sky Survey Alternate Data Release 1 (sample size=369; S. Manik et al. 2025), so our results are in accordance with those reported in the literature.

Columns 8 and 9 of the Table 3 give the number of sources showing arm-length ratio≥2 and the corresponding fraction for each radio AGN morphological class. We find that: (1) the cluster association fractions and large arm-length asymmetry fractions are comparable for the I and I/II classes, (2) the cluster association fraction is about 2–4 times larger than the arm-length asymmetry fraction for the type-II class. For a full sample with secure+possible identifications, fraction of sources showing arm-length asymmetry is ∼18%, slightly smaller than cluster association fraction of ∼28%, and (3) WATs, NATs and HTs classes show a significantly larger cluster association fraction (∼35%–∼60%) than arm-length asymmetry fraction (∼12%–∼30%), and (3) the number of sources with DD, X and Z-shaped morphologies are too few to make any concluding remark about the relation between the arm-length asymmetry and their respective cluster environments.

Taking into account the cluster association and arm-length asymmetry fractions, the similar fractions found for the I and I/II classes and different fractions for the II class can be explained within the framework of stan-dard models of FR jets (C. M. Urry & P. Padovani 1995). According to these, the jets of type-I and type-II sources, although starting super-sonically with high Mach numbers (≥1), exhibit different deceleration and collimation properties as they propagate in the intracluster medium. The jets of type-I sources decelerate on a kiloparsec scale due to instabilities in the flow(e.g., O. Porth & S. S. Komissarov 2015) or mass loading by stellar winds (e.g., M. Perucho et al. 2014) and become transonic, losing their collimation halfway between the host galaxy and the tip of their lobe(s). Such jets will be susceptible to disruption when propagating in a relatively dense cluster medium. The jets of type-II sources, on the other hand, maintain their collimation all the way to the tip of their lobes and are still highly supersonic as they terminate in a compact hot spot and, therefore, would be less susceptible to disruption.

WATs and NATs sources show significantly higher cluster association than arm-length asymmetry fractions. Since the vast majority of them are found in merging cluster environments (Section 4.2), we speculate that any arm-length asymmetries in WATs and NATs sources are smoothed out by the turbulent intracluster medium conditions, for example, such as gas sloshing (lobes of a WAT source are oriented along the slohing spiral in Abell 1763; E. M. Douglass et al. 2018), or shock fronts (lobes of NAT source in the southern subcluster of Abell 1569 are bent due to the shock created after the subcluster merger; J. Tiwari & K. P. Singh 2022).

### 4.4. *Evolutionary tracks of FR II sources: Power-linear size diagram*

A key diagnostic tool for understanding the evolution and physical processes in radio-loud AGN in general is the radio power (P) versus linear size (D) diagram, commonly denoted as the P–D diagram. In this diagram, the monochromatic radio luminosity is plotted against the projected linear size of the radio source, spanning from compact sources with sizes ≤1-10 kpc to very extended giants up to several Mpc. The radio source on this plot follows a very specific evolutionary track, which depends on its properties such as jet kinetic power or the environment density (C. R. Kaiser et al. 1997; J. Machalski et al. 2004; M. J. Hardcastle 2018; R. J. Turner et al. 2023). In order to draw the P-D diagram of FR type-II sources in our sample, we adopted the analytical model of source evolution by C. R. Kaiser & P. Alexander (1997); C. R. Kaiser et al. (1997). Within the framework of this dynamical evolutionary model, the jets' kinetic power ($Q_0$) remains constant during the source lifetime as they propagate in the medium surrounding the host galaxy, inflating a bubble of low-density, high-pressure materials that contains relativistic electrons that generate radio emission as they interact with the ambient magnetic field. The jets terminate in the IGM, where they form jet shocks, and their thrust (i.e., the ram pressure of the material) is distributed over the working surface, or hotspot. Most of it is transported to the hotspot, while the excess energy is stored in the cocoon, which is inflated around the jet. This, in turn, drives the bow shock into the surrounding (gas) medium. In such a scenario, the radio source expands in a "self-similar" fashion where the lengthening of the cocoon is governed by the thrust employed by the advancing hotspot (balanced by the ram pressure of the IGM). In contrast, its width is governed by the internal pressure of the cocoon. Twin jets deliver energy equal to $2\,Q_0 t$ during the source lifetime, $t$, and the particles lose energy via adiabatic and radiative losses. The impact on the lifetime of FR type-II sources at high redshift will result in higher cooling due to increased inverse-Compton cooling losses. This is a fully analytical model in which the geometry and internal pressure of the expanding lobe are related to the jet dynamics.

Figure 5 shows the P-D diagram for our sample of FR type-II sources drawn for jet energies $10^{36}$, $10^{37}$, $10^{38}$, and $10^{39}$ ergs. The maximum redshift of the type-II sources is 0.52 with a peak around 0.2; therefore, we choose the redshifts as 0.01, 0.2, 0.4, and 0.6. The source ages are indicated at an interval of 1 Myr between ages of 1 Myr to 10 Myr, subsequently increasing to 10 Myr for ages between 10 Myr to 100 Myr and 100 Myr for ages between 100 Myr to 1 Gyr. For our sources, we compute the $K$-corrected rest-frame 1.4 GHz luminosities from the observed luminosities using the relation $L_{rest,total} = (1+z)^{\alpha-1} L_{obs,total}$. We use a constant $\alpha$=0.7, typical for radio AGN (R. Kondapally et al. 2022), as it is expected that the total luminosity of FR II sources is dominated by steep spectrum lobes, rather than a flat spectrum radio core (see also, P. Dabhade et al. 2020b). We note that ∼80% of our type-II sources show a total flux to core flux ratio larger than 3; therefore, scaling the observed total radio luminosity to the rest-frame luminosity with one spectral index corresponding to the steeper-spectrum lobe plasma is justified. The positions of our FR type-II AGNs on this diagram show that the vast majority of these sources have jet kinetic powers in the range $10^{36-38}$ erg. The position of the time-evolution markers on the tracks denoted by the plus symbols shows that all type-II sources have relatively long lifetimes (2 Myr to 100 Myr) and we expect them to reach real sizes of a few kpc to 10 Mpc before falling below the typical detection limits of the FIRST and NVSS surveys.

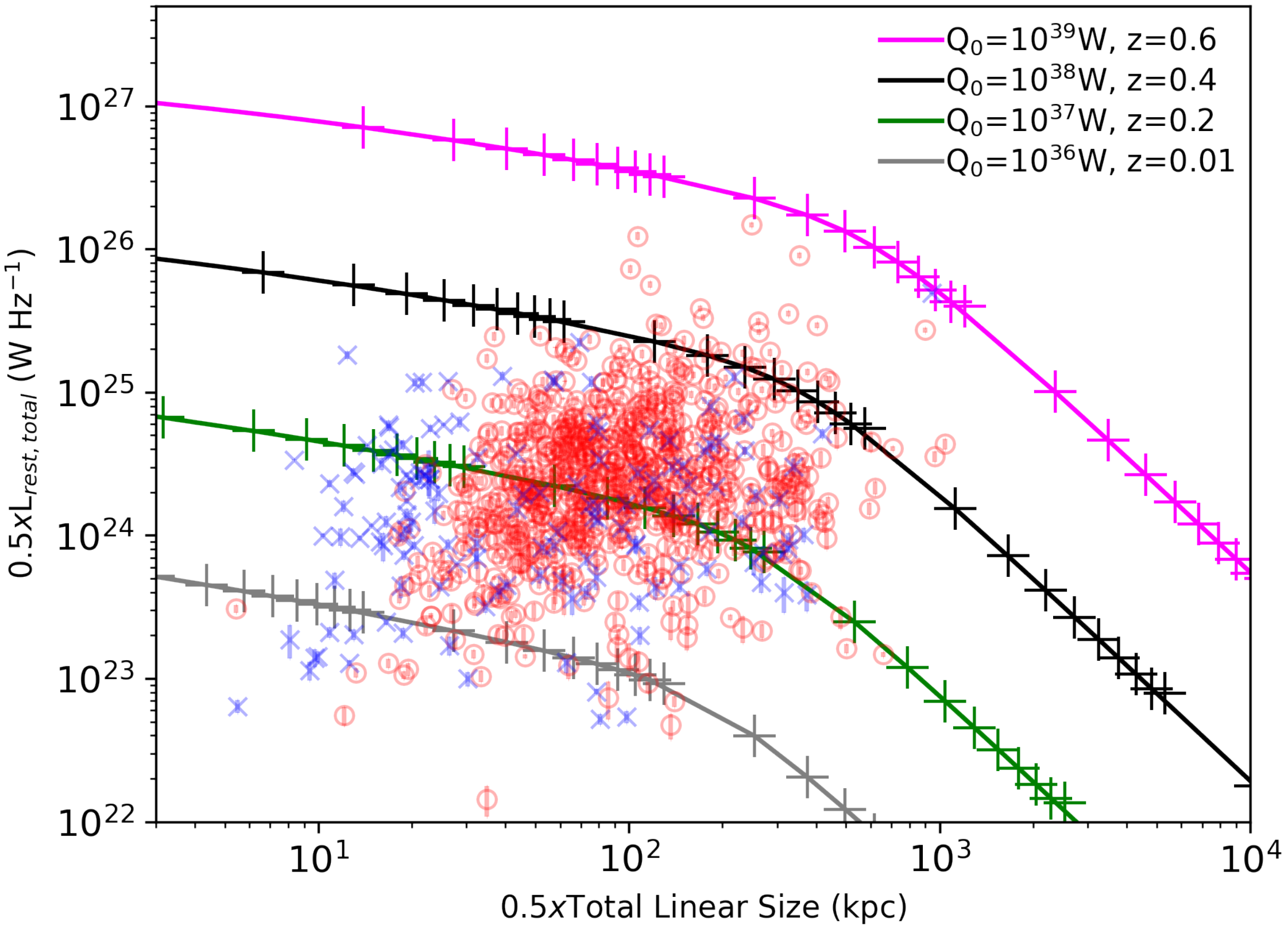

**Figure 5.** P-D diagram for the FR II sources using the analytical model of (C. R. Kaiser et al. 1997). The evolutionary tracks are shown for jet powers $Q_0{=}\times10^{39}$ W at z=0.6, $Q_0{=}\times10^{38}$ W at z=0.4, $Q_0{=}\times10^{37}$ W at z=0.2 and $Q_0{=}\times10^{36}$ W at z=0.01. The axial ratio is 2, and the jet material is composed of cold electron-positron plasma with adiabatic index=5/3. Both the cocoon and the magnetic field energy densities have a relativistic equation of state (i.e., adiabatic index =4/3). The adiabatic index of jet is 5/3. The density of the ambient gas is $7.2\times10^{-22}$ kg $m^{-3}$ and the core radius of the cluster is 2 kpc. The total radiated power is computed for electron Lorentz factors $10^3$ to $10^6$. All the constants used in deriving radio power and size are computed according to those given in the (C. R. Kaiser & P. Alexander 1997; C. R. Kaiser et al. 1997). The tracks are plotted for half radio-power (total luminosity/2) and half jet-length (total linear size/2). The source ages (shown by a plus sign on the tracks) are indicated at an interval of 1 Myr between ages of 1Myr to 10 Myr, subsequently increasing to 10 Myr for ages between 10Myr to 100 Myr, and 100 Myr for ages between 100Myr to 1 Gyr.

For comparison, the P-D diagram for the 3rd revised Cambridge catalog (3CRR; R. A. Laing et al. 1983) sources presented in Figure 1 of C. R. Kaiser et al. (1997) indicates that they are dominated by relatively stronger type-II sources with jet energy $\geq 10^{39}$ ergs, in contrast, to the type-II sources presented here where the majority of sources show jet energies $\leq 10^{38}$ ergs. This is because the 3CRR survey had relatively poor sensitivity (10 Jy) compared to the radio catalogs used here. In particular, the P-D tracks of the LOFAR sources also show that they are dominated by radio AGNs with jet energies $\leq 10^{38}$ ergs (Figure 8; M. J. Hardcastle et al. 2019); this is because LoTSS DR 1 catalog has sensitivity limit of $\sim$ 1mJy, comparable to ROGUE I catalog. We note that the self-similar jet evolution model of C. R. Kaiser & P. Alexander (1997) is a simple model in which the source expands as a cylinder and the jet power remains constant throughout its lifetime. More realistic models that predict the evolutionary tracks of FR II sources that include a remnant phase of AGN where the jet activity stops after some time exist (E. Kuligowska 2017; R. J. Turner et al. 2023). Finally, more complex semi-analytical and numerical models by M. J. Hardcastle et al. (2019) also give the P-D tracks of FR sources, yet the C. R. Kaiser et al. (1997) is relevant for predicting the jet length with source age, which gives the same dependence when semi-analytical models are considered (see, Table 1 of R. J. Turner & S. S. Shabala 2023).

**Table 4**. Median values of the distributions of the radio AGN parameters for a given morphological class. Values for 'secure' identifications are followed by values for 'possible' identifications and for all sources ('secure+possible' identifications) in the next lines. Columns are: (1) name of the radio AGN morphological class, (2) redshift, (3) total angular size, (4) total projected linear size, (5) 1.4 GHz radio core luminosity, (6) 1.4 GHz radio total luminosity, (7) black hole mass (8) $r'$-band absolute optical magnitude, and (9) stellar mass

| Class | z | Ang. Size ($''$) | Li. Size (kpc) | $L_{\rm core}$ ($\times10^{23}$W Hz$^{-1}$) | $L_{\rm total}$ ($\times10^{24}$W Hz$^{-1}$) | $M_{BH}$ ($\times10^{8}M_{\odot}$) | $M'_r$ (mag) | $M_\star$ ($\times10^{11}M_{\odot}$) |
|---|---|---|---|---|---|---|---|---|
| (1) | (2) | (3) | (4) | (5) | (6) | (7) | (8) | (9) |
| I | 0.1870 | 44 | 142.1 | 9.6 | 4.4 | 3.7 | -22.80 | 4.6 |
| pI | 0.2169 | 25 | 97.5 | 14.8 | 2.9 | 3.8 | -22.77 | 4.5 |
| I(all) | 0.1973 | 38 | 122.4 | 11.4 | 3.6 | 3.8 | -22.79 | 4.6 |
| II | 0.2103 | 50 | 180.2 | 6.9 | 5.1 | 3.5 | -22.79 | 4.5 |
| pII | 0.2251 | 25 | 92.2 | 7.5 | 3.1 | 3.8 | -22.75 | 4.4 |
| II(all) | 0.2131 | 48 | 162.3 | 7.0 | 4.7 | 3.6 | -22.79 | 4.5 |
| I/II | 0.1953 | 51 | 165.7 | 7.1 | 4.6 | 3.6 | -22.71 | 4.3 |
| pI/II | 0.2040 | 45 | 160.0 | 8.2 | 3.1 | 3.4 | -22.76 | 4.1 |
| I/II(all) | 0.1988 | 49 | 163.0 | 7.5 | 4.0 | 3.5 | -22.74 | 4.2 |
| WAT | 0.2148 | 48 | 175.4 | 11.1 | 6.1 | 4.1 | -22.75 | 4.7 |
| pWAT | 0.1799 | 67 | 245.9 | 6.4 | 3.3 | 3.3 | -22.79 | 4.3 |
| WAT(all) | 0.2101 | 50 | 183.5 | 10.1 | 5.4 | 4.0 | -22.75 | 4.6 |
| NAT | 0.2098 | 42 | 135.6 | 11.5 | 6.4 | 3.9 | -22.67 | 4.2 |
| pNAT | 0.2050 | 33 | 109.8 | 12.0 | 4.5 | 3.4 | -22.55 | 3.8 |
| NAT(all) | 0.2058 | 39 | 134.5 | 11.8 | 6.2 | 3.7 | -22.63 | 4.2 |

The large sample of radio AGNs with different morphological classifications in the ROGUE I catalog allows us to compare the physical properties across classes. For comparison of physical properties across different AGN morphological classes, we use the I, II, I/II, WAT, and NAT categories, as they have more than 100 sources identified with a 'secure' classification, to enable meaningful statistical inference. Figure 6 gives the distributions, for $z$ (panels a1-a5), angular size (panels b1-b5), linear size (panel c1-c5), $L_{\rm core}$ (panels d1-d5), $L_{\rm total}$ (panels e1-e5), $M_{BH}$ (panels f1-f5), $M_r$ (panels g1-g5), and $M_\star$ (panels h1-h5) for these morphological classes. Table 4 gives the median values of these distributions. We find that the type-I sources are at slightly lower redshifts (median $z$ $\sim$0.1973) than the other classes. The median values of angular sizes of type-I sources are comparable to the NAT sources ($\sim$39$''$), and smaller than the rest of the classes ($\sim$49$''$). The median values of the total projected linear sizes are slightly smaller for the type-I class ($\sim$122 kpc) and NAT class ($\sim$135 kpc), than those of type-II ($\sim$163 kpc), type-I/II ($\sim$ 163 kpc), and WAT ($\sim$183 kpc) classes. The median values of $L_{\rm core}$ show a range 7.0$\times$10$^{23}$–11.8$\times$10$^{23}$ W Hz$^{-1}$ for the considered classes. The median values of $L_{\rm total}$ show a range 3.6$\times$10$^{24}$–6.2$\times$10$^{24}$ W Hz$^{-1}$ for the considered classes. The median values of black hole and stellar masses range 3.5$\times$10$^{8}$–4.0$\times$10$^{8}$ $M_\odot$ and 4.2$\times$10$^{11}$–4.5$\times$10$^{11}$ $M_\odot$, respectively, for the considered classes. The median optical r-band magnitudes range from -22.63 to -22.79 for the considered classes.

Next, we compare these properties across the classes, and the results of the two-sample KS test are given in Table 5. The null-hypothesis that the two distributions are drawn from the same parent distribution is rejected for $p-$values $\leq$0.05 (shown by the blue color) and $\leq$0.003 (indicated by the red color), corresponding to confidence levels of 95% and 99.7%, respectively. We note that redshift distributions differ at a 95% confidence level between the type-I and type-II classes, while remaining insignificant between other classes. The total angular size distributions are significantly different between FR type-I and all classes, whether considering the sample of sources with secure identifications only or all sources with secure+possible identifications. The angular size distributions are different between NAT and all other classes. The projected linear size distributions are also significantly different across all classes, whether considering the sample of sources with secure identifications only or all sources with secure+possible identifications. The radio core luminosity distributions are different for pairs, comparing: type-I and type-II, type-II and WAT, type-II and NAT, and I/II and NAT. The distributions of total radio luminosity are found to be different between pairs: type-I and type-II, type-I and WAT, type-I and NAT, type-II and NAT, I/II and WAT, and I/II and NAT classes. The distributions of $M_{\rm BH}$ are statistically indistinguishable between the classes. The distributions of optical luminosity of the host are found to differ at the 95% confidence level between type-II and the NAT classes, considering secure+possible identifications, while remaining indistinguishable between the other pairs. The distributions of galaxy mass, $M_\star$, are also found to be statistically indistinguishable between the classes.

Since the radio AGN classes are situated at similar redshifts, apart from one exception of type-I and type-II, it is reasonable to state that, on average, these sources are situated at similar distances. Therefore, the Malmquist bias should not affect the inter-comparison of physical properties, derived from luminosities, between the classes. The sources have similar $M_\star$ and $M_{\rm BH}$ distributions, yet they exhibit significant differences in the distributions of total angular/projected linear sizes across some classes. Moreover, differences in linear size between the classes are not directly reflected in the distributions of total luminosity, indicating the jet power (C. J. Willott et al. 1999). It means that indeed the linear sizes attained by extended radio AGNs cannot be attributed directly to a single parameter (i.e, jet power) but instead other parameters such as host galaxy mass, environment, or accretion efficiency play a significant role in shaping them (K. M. Blundell et al. 1999; P. N. Best & T. M. Heckman 2012; J. H. Croston et al. 2019). We can scrutinize our findings for the type-I and type-II classes as they differ in total luminosities and total angular/projected linear sizes, but exhibit similar distributions when $M_\star$ is considered. Suppose if sources with high power jets (e.g., type-II) are expected to attain large linear sizes compared to low power jets (e.g., type-I), then one should expect different distributions in $M_\star$ and $M_r$ for the host galaxy between these classes if the Fanaroff-Riley type-I/type-II dichotomy is to be believed (F. N. Owen & M. J. Ledlow 1994; M. J. Ledlow & F. N. Owen 1996). According to these authors, for a given mass of elliptical host galaxy, there is a unique transition jet flux such that for higher ambient pressures, the type-I/type-II jet transition occurs at a higher jet energy flux. However, the type-I/type-II dichotomy does not necessarily imply that the projected linear sizes attained by the sources would be different, as recent models of jet evolution for FR sources show radio power–linear size tracks, depending not only on the jet kinetic power, source age, but also on the density profile of the medium (see, R. J. Turner & S. S. Shabala 2023). Therefore, it is difficult to summarize our findings simply by comparing a few observables for the FR class.

Next, similar comparisons between bent-angle sources and straight-line FR class AGNs would be difficult, as bent-angle sources are predominantly found in cluster environments, where the density profile can be different depending on their location relative to the cluster core (see, Section 4.2). Moreover, ram pressure exerted on the lobes by local intracluster medium conditions can also alter the arm lengths. In our follow-up study, we aim to examine these and other properties, such as accretion rate, high/low excitation classification, distribution of optical light in the host galaxy, galaxy extinction, and age of the stellar population, to understand what causes the sources to attain large linear sizes.

**Table 5.** p-values of two-sample KS-test performed using physical properties for different AGN classes. Values for 'secure' identifications are followed by values for 'secure+possible' identifications in parentheses. p-values ≤0.003 (shown in red color) and ≤0.05 (shown in blue color) rejects the null-hypothesis that the two classes are drawn from one parent population at a confidence levels ≥99.7% and 95%, respectively.

| Class I \ Class II | I | II | I/II | WAT | NAT |
|---|---|---|---|---|---|
| — z — | | | | | |
| I | – | 0.014(0.047) | 0.132(0.176) | 0.016(0.117) | 0.550(0.864) |
| II | – | – | 0.305(0.094) | 0.514(0.987) | 0.217(0.412) |
| I/II | – | – | – | 0.215(0.289) | 0.685(0.703) |
| WAT | – | – | – | – | 0.111(0.304) |
| — Total Angular Size — | | | | | |
| I | – | 0.322($< 10^{-3}$) | 0.604(0.007) | 0.621($< 10^{-4}$) | 0.194(0.037) |
| II | – | – | 0.593(0.568) | 0.646(0.072) | 0.014(0.009) |
| I/II | – | – | – | 0.542(0.065) | 0.026(0.001) |
| WAT | – | – | – | – | 0.061(0.004) |
| —Projected Total Linear Size — | | | | | |
| I | – | (0.002$<10^{-5}$) | 0.236(0.003) | 0.001($<10^{-8}$) | 0.030(0.020) |
| II | – | – | 0.353(0.495) | 0.285(0.045) | $<10^{-4}$($<10^{-4}$) |
| I/II | – | – | – | 0.127(0.014) | 0.003(0.001) |
| WAT | – | – | – | – | $<10^{-3}$($<10^{-5}$) |
| — $L_{core}$ — | | | | | |
| I | – | 0.003($<10^{-5}$) | 0.360(0.133) | 0.437(0.540) | 0.454(0.842) |
| II | – | – | 0.371(0.194) | $<10^{-4}$($<10^{-3}$) | 0.005(0.001) |
| I/II | – | – | – | 0.054(0.150) | 0.044(0.056) |
| WAT | – | – | – | – | 0.963(0.710) |
| —$L_{total}$ — | | | | | |
| I | – | 0.144(0.003) | 0.368(0.283) | 0.008($<10^{-3}$) | 0.013(0.002) |
| II | – | – | 0.909(0.230) | 0.184(0.170) | 0.072(0.048) |
| I/II | – | – | – | 0.244(0.018) | 0.082(0.006) |
| WAT | – | – | – | – | 0.725( 0.738) |
| — $M_{BH}$ — | | | | | |
| I | – | 0.496(0.450) | 0.632(0.489) | 0.364(0.536) | 0.993(0.971) |
| II | – | – | 0.972(0.956) | 0.044(0.190) | 0.317(0.502) |
| I/II | – | – | – | 0.476(0.235) | 0.442(0.566) |
| WAT | – | – | – | – | 0.573(0.447) |
| — $M_r$ — | | | | | |
| I | – | 0.989(0.766) | 0.407(0.419) | 0.741(0.813) | 0.130(0.059) |
| II | – | – | 0.304(0.234) | 0.750(0.786) | 0.127(0.032) |
| I/II | – | – | – | 0.854(0.622) | 0.878(0.343) |
| WAT | – | – | – | – | 0.878(0.343) |
| — $M_\star$ — | | | | | |
| I | – | 0.415(0.322) | 0.750(0.458) | 0.658(0.762) | 0.603(0.414) |
| II | – | – | 0.418(0.223) | 0.216(0.183) | 0.644(0.320) |
| I/II | – | – | – | 0.517(0.298) | 0.900(0.998) |
| WAT | – | – | – | – | 0.573(0.459) |

## 5. SUMMARY

We present hand-curated size measurements for a sample of 2,002 multiple-component radio AGNs belonging to Fanaroff-Riley type-I, -II, and -I/II classes, as well as double-double, X-shape, Z-shape, wide-angle tail, narrow-angle tail, and head-tail radio galaxies, identified in the ROGUE I catalog. We also study their large-scale environments, arm-length ratios, and provide an inter-comparison of projected total linear sizes, core and total radio luminosity, absolute optical magnitude, black hole mass, and stellar mass between the classes. The redshift range of galaxies studied here is between 0.01 and 0.54. Below, we summarize our main findings:

1. We give the most extensive catalog of total angular and projected linear sizes for a sample of 2,002 radio AGNs. The distributions of total angular and projected linear sizes are similar for the 'secure' and 'possible' identifications of radio morphology, indicating that no significant bias has entered either into morphology assignment or size measurements of the radio sources.

2. The range of total angular sizes is between $\sim5''$ and $\sim1{,}100''$, translating to the projected linear sizes between ~10 kpc to ~2,200 kpc for the redshift range of radio AGNs in the ROGUE I catalog.

The majority of sources exhibit total angular sizes $\leq 100''$ and projected total linear sizes $\leq$200 kpc.

3. 211 ($\sim$ 10%) radio AGNs present small linear sizes $\leq$60 kpc, and 71 (3.4%) sources present large linear sizes corresponding to giant classification $\geq$700 kpc.

4. We find that total angular size is approximately anti-correlated with redshift with an index of fitted power-law $\approx -0.7$. The projected total linear size is not correlated with $z$. The 1.4 GHz total flux density is moderately correlated with angular size and varies as angular size to the power of $\sim$0.6. The 1.4 GHz total radio power correlates with the projected linear size, with a fitted power-law index $\approx$ 0.5. For our volume-limited sample, the index between radio power and linear size is flatter, $\sim$0.2, compared to the full sample, indicating that Malmquist bias is significantly affecting this trend. The trends seen in angular/linear size vs. $z$ and flux density vs. $z$ are not affected by the Malmquist bias in our data. The result of the 2D power-fit of projected linear size vs. redshift and total radio power indicates that smaller linear size sources are also radio-weak.

5. Approximately 34% of multiple-component radio AGNs are found to be associated with galaxy clusters. A much larger proportion than the 10-20% detected using the LoTSS DR 1 and DR 2 data.

6. We find that $\sim$50%, of bent-angle radio AGNs (WAT, NAT, and HT) compared to $\sim$30% of straight-line radio AGNs (type-I, type-II, and type-I/II) reside in galaxy cluster environments. The cluster association fraction of bent-angle and straight-line FR sources differs at $\geq$95% confidence levels, indicating that bent-angle sources are more likely to reside in a cluster environment.

7. About 15.8% of our sample of two-sided radio AGNs exhibit arm-length asymmetry ratio $\geq$2.

8. We find that mean galaxy number densities of clusters that inhabit the radio AGNs of Fanaroff-Riley type-I, type-II, type-I/II, WAT, NAT, and HT classes are comparable. It indicates that local ICM conditions play an important role in determining the bent morphologies of radio AGNs.

9. We compared the fractions of sources showing large arm-length asymmetries to the galaxy cluster association for each radio AGN morphological class. We find that the fractions of sources inhabiting cluster environments and exhibiting large lobe-length asymmetry are strikingly similar for the Fanaroff-Riley type-I and I/II classes ($\sim$20-30%). In contrast, the Fanaroff-Riley type-II class show much higher cluster association fraction ($\sim$30%) compared to arm-length asymmetry fraction ($\sim$18%). A large fraction of bent-angle sources, WATs ($\sim$43%) and NATs ($\sim$58%) are found to be associated with cluster environments; however, a much smaller fraction is showing large arm-length asymmetry ($\sim$11-30%).

10. The evolutionary tracks of the FR II sources obtained from the analytical model of (C. R. Kaiser et al. 1997) indicate AGNs with jet kinetic energies $\leq 10^{38}$ erg s$^{-1}$ dominate the ROGUE I catalog and comprise of both young ($\leq$10 Myr) and old ($\leq$100 Myr) AGNs.

11. We find that the type-I sources are at slightly lower redshifts (median $z$ $\sim$0.1973) than the other classes. The median values of angular sizes of type-I sources are comparable to the NAT sources ($\sim 39''$), and smaller than the rest of the classes ($\sim 49''$). The median values of the total projected linear sizes are slightly smaller for the type-I class ($\sim$122 kpc) and NAT class ($\sim$135 kpc), than those of type-II ($\sim$163 kpc), type-I/II ($\sim$ 163 kpc), and WAT ($\sim$183 kpc) classes. The median values of $L_{\rm core}$ show a range $7.0\times10^{23}$–$11.8\times10^{23}$ W Hz$^{-1}$ for the considered classes. The median values of $L_{\rm total}$ show a range $3.6\times10^{24}$–$6.2\times10^{24}$ W Hz$^{-1}$ for the considered classes. The median values of black hole and stellar masses range $3.5\times10^{8}$–$4.0\times10^{8}$ $M_\odot$ and $4.2\times10^{11}$–$4.5\times10^{11}$ $M_\odot$, respectively, for the considered classes. The median optical r-band magnitudes range from -22.63 to -22.79 for the considered classes.

12. The large number of sources identified with different radio AGN morphologies in the ROGUE I catalog allowed us to compare a few astrophysical properties within the classes. In particular, we compared the redshifts, projected total angular and linear sizes, core and total radio luminosities, optical luminosities, stellar masses, and BH masses between the Fanaroff-Riley type-I, type-II, type-I/II, WAT, and NAT classes. Given that almost all radio AGNs ($\geq$99%; column 7 of Table 2) are inhabited by elliptical hosts in our sample, the significant differences found in the distributions of total angular/projected linear sizes between the radio AGN classes, which are not reflected either in distributions of redshift, total radio luminosity, optical luminosity; we are yet to answer the main question as to what makes the radio AGNs attain specific linear sizes.

## ACKNOWLEDGMENTS

We are thankful to the reviewer for suggestions to clarify the paper and to expand on some of our results. The publication has been supported by a grant from the Faculty (Faculty of Physics, Astronomy and Applied Computer Science) under the Strategic Programme Excellence Initiative at Jagiellonian University. AW was supported by the GACR grant 21-13491X. S.D. acknowledges support from the Polish National Science Center through project UMO-2023/51/D/ST9/00147. US acknowledges support from the National Research Foundation of South Africa. NVA acknowledges support from Conselho Nacional de Desenvolvimento Científico e Tecnológico (CNPq). This study was financed in part by the Coordenação de Aperfeiçoamento de Pessoal de Nível Superior - Brasil (CAPES) – Finance Code 001. AW, GS, and NVA acknowledge the research project no. 2021/43/B/ST9/03246 financed by the National Science Centre/Narodowe Centrum Nauki. We thank Alexander Lenart, Pratik Dabhade, Dorota Kozieł-Wierzbowska, Krzysztof Chyży, and Dominika Hunik for discussions.

Funding for the Sloan Digital Sky Survey V has been provided by the Alfred P. Sloan Foundation, the Heising-Simons Foundation, the National Science Foundation, and the Participating Institutions. SDSS acknowledges support and resources from the Center for High-Performance Computing at the University of Utah. SDSS telescopes are located at Apache Point Observatory, funded by the Astrophysical Research Consortium and operated by New Mexico State University, and at Las Campanas Observatory, operated by the Carnegie Institution for Science. The SDSS web site is www.sdss.org.

SDSS is managed by the Astrophysical Research Consortium for the Participating Institutions of the SDSS Collaboration, including the Carnegie Institution for Science, Chilean National Time Allocation Committee (CNTAC) ratified researchers, Caltech, the Gotham Participation Group, Harvard University, Heidelberg University, The Flatiron Institute, The Johns Hopkins University, L'Ecole polytechnique fédérale de Lausanne (EPFL), Leibniz-Institut für Astrophysik Potsdam (AIP), Max-Planck-Institut für Astronomie (MPIA Heidelberg), Max-Planck-Institut für Extraterrestrische Physik (MPE), Nanjing University, National Astronomical Observatories of China (NAOC), New Mexico State University, The Ohio State University, Pennsylvania State University, Smithsonian Astrophysical Observatory, Space Telescope Science Institute (STScI), the Stellar Astrophysics Participation Group, Universidad Nacional Autónoma de México, University of Arizona, University of Colorado Boulder, University of Illinois at Urbana-Champaign, University of Toronto, University of Utah, University of Virginia, Yale University, and Yunnan University.

The National Radio Astronomy Observatory is a facility of the National Science Foundation operated under cooperative agreement by Associated Universities, Inc. CIRADA is funded by a grant from the Canada Foundation for Innovation 2017 Innovation Fund (Project 35999), as well as by the Provinces of Ontario, British Columbia, Alberta, Manitoba and Quebec.

## AUTHOR CONTRIBUTIONS

AG is responsible for research concept, measurement of linear sizes, derivation of P-D tracks using the C. R. Kaiser et al. (1997) model, comparison of properties, manuscript writing, and submitting. AG cross-validated the size measurements for the entire catalog, while AM performed the cross-validation of sizes for a part of the catalog. AM and SD measured the fraction of sources in clusters. US contributed to the discussion on Malmquist bias and the computation of galaxy number density in clusters. MS contributed to the realization of the P-D tracks. AW, GS, and NVA performed the editorial corrections to the paper. SN contributed to the discussion on P-D tracks. All co-authors have contributed to discussions on the results.

*Facilities:* PO:1.2m, UKST, Sloan, VLA

*Software:* Numpy(C. R. Harris et al. 2020), Scipy(P. Virtanen et al. 2020), Pandas(W. McKinney 2010), Matplotlib(J. D. Hunter 2007), Astropy( Astropy Collaboration et al. 2013, 2018, 2022)

## APPENDIX

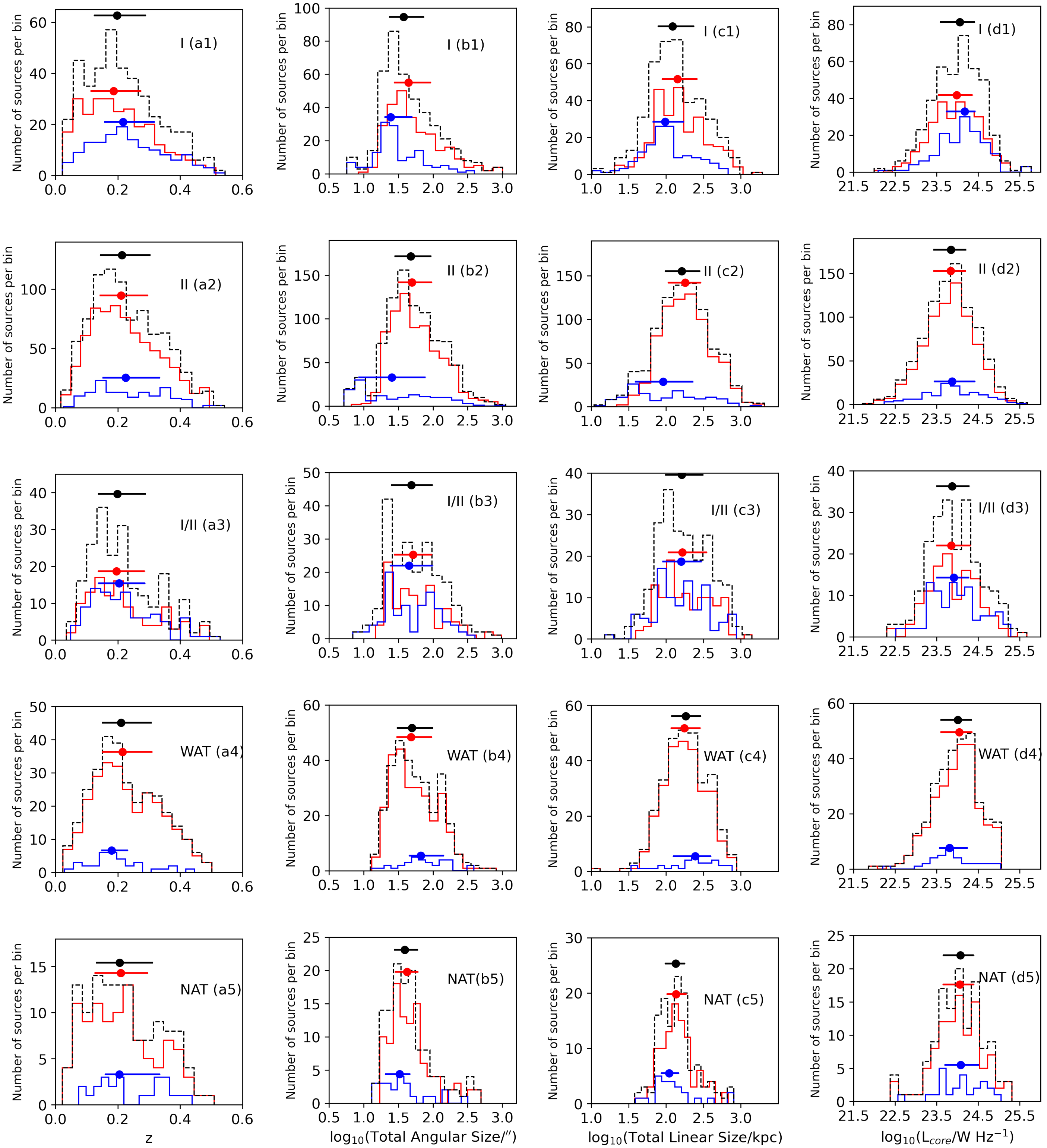

**Figure 6.** Histograms showing the distributions of the parameters of radio AGN morphological classes (Table 5). The solid lines in red and blue and dashed line in black give distributions considering 'secure', 'possible', and 'all' (secure+possible) identifications. The horizontal segments on top of the histograms indicate the 16th and 84th percentiles, with a filled circle marking the 50th percentile. The names of the morphological classes are given inside the panel.

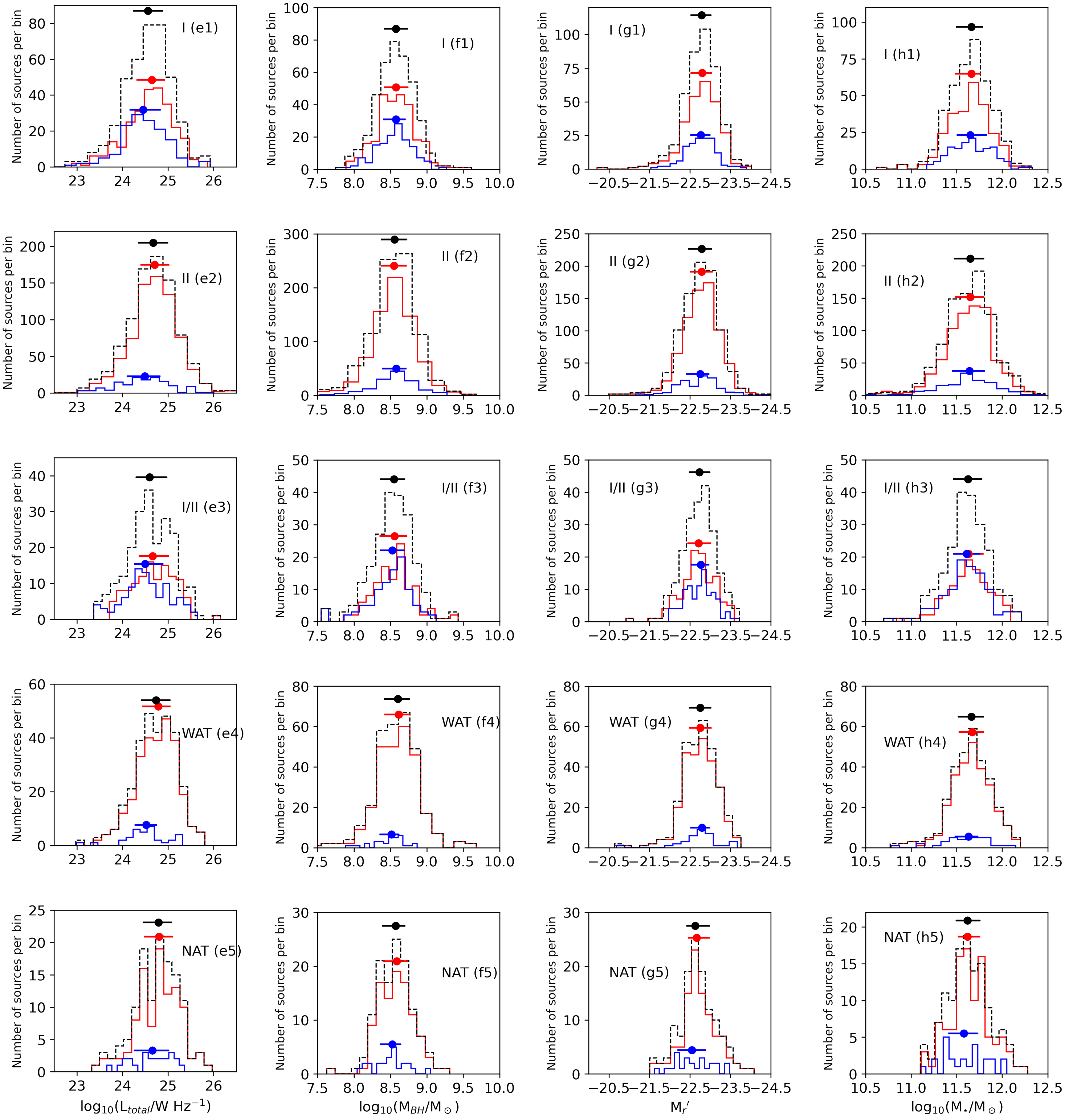

**Figure 6 (Cont.).**

## A. EXAMINING THE 81 'FDB' FLAGGED SOURCES

Since we measured the sizes of these sources using the two radio components detected in FIRST maps, with one component falling $\leq 3''$ from the optical center (our definition of radio core), we further examined these sources using higher than FIRST resolution radio maps, with the expectation of detecting a radio core component. For this purpose, we use the publicly available VLASS Quick Look (QL) images[11], made at a central frequency of 3.0 GHz with an angular resolution of $2.5''$(Y. A. Gordon et al. 2020). The astrometric accuracy of sources detected in VLASS QL images is $\sim 1.0''$ (Y. A. Gordon et al. 2021). 65 out of 81 sources present a radio component within $30''$ radius of the optical center (Y. A. Gordon et al. 2021), while the remaining 16 'FDB' flagged sources are not detected in VLASS QL images. In Figure A.1, we give the VLASS and FIRST contour maps of these sources overlayed on optical DSS images. Sources with ROGUE I IDs 445, 939, 1416, 2678, 2843, 3012, 4681, 5995, 11067, 11328, 13230, 13593, 13772, 23148, 23774, 25199, 23770, 28129, 28318, 29499, 30190, and 31965 present a clear radio core $\leq 1.5''$ from the optical center, closer than presented by the FIRST map. We do not find a clear VLASS core in the remaining maps, where the radio lobes are often detected near those in the FIRST maps. For many of our sources, the VLASS maps reveal the presence of a lobe component, farther than the FIRST lobe (ROGUE IDs: 2678, 4681, 5995, 13772, 18781, 25199, 27528, 29499, and 29942). Since 'FDB' flagged sources are only 4% of the total sample of 2,002 sources analysed here, we do not expect the source statistics to change, such as the size estimations (by not considering the VLASS-detected component farther than the FIRST-detected component from the optical host galaxy), or the core fluxes (which should be replaced with flux of the VLASS-detected component), by their inclusion in our sample. Therefore, to maintain uniformity with the analysis procedure, we keep the size measured on the FIRST maps and the core fluxes from the ROGUE I catalog. For the case-by-case study, we caution the readers to carefully examine the 'FDB' flagged sources.

[11] https://www.cadc-ccda.hia-iha.nrc-cnrc.gc.ca/en/vlass/

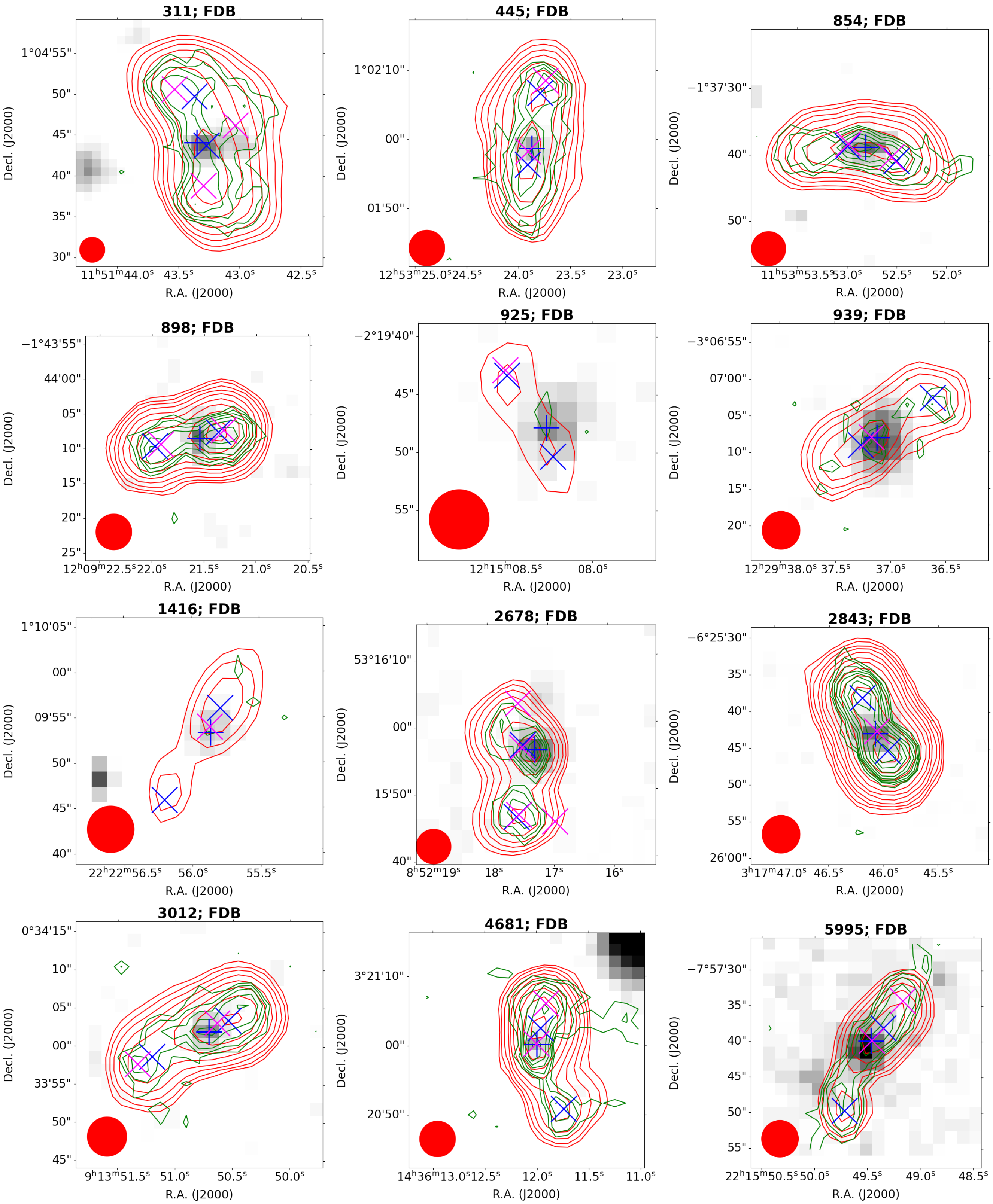

**Figure A.1.** Radio-optical overlay map of the galaxies flagged as 'FDB' in the method of size measurement column of Table 2. The FIRST and VLASS intensities are shown in red and green contours, respectively, while the DSS intensity is shown in grayscale. The radio intensities are drawn at typical $3\sigma$ sensitivity limits of these surveys (0.2 mJy beam$^{-1}$ and 120 $\mu$Jy beam$^{-1}$ for FIRST and VLASS, respectively) and increase by $(\sqrt{2})^n$ where $n$ range from 0,1,2,..,20. A blue plus sign marks the optical position of the host galaxy from the SDSS survey, while the blue and magenta cross symbols mark the positions of the lobe components from the FIRST and the VLASS surveys, respectively. The ROGUE I ID of galaxy and the keyword describing the method of size measurement (column 16; Table 1) is given at the top of each panel, while the FIRST beam, shown with a red filled circle, is indicated at the bottom left corner.

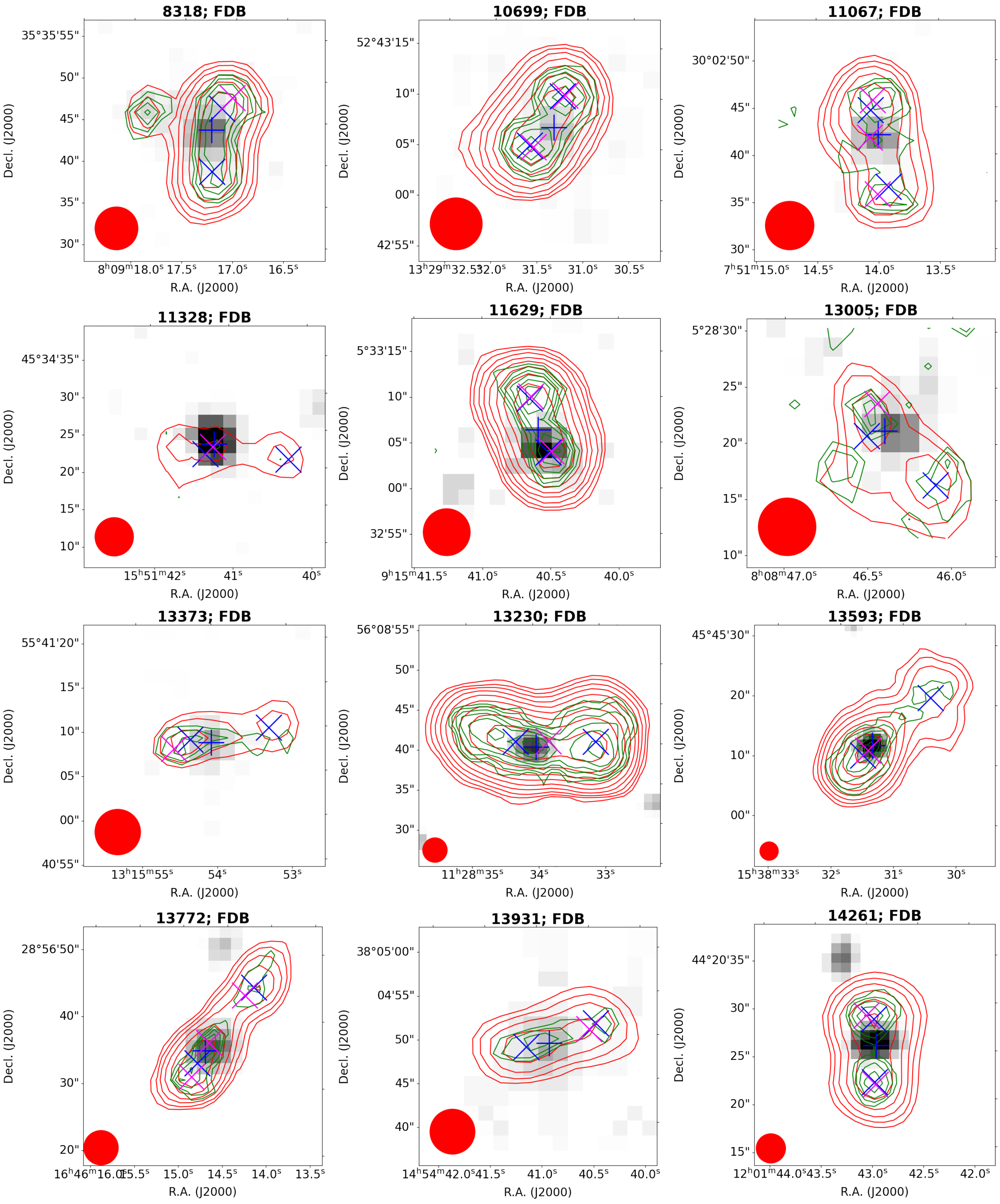

**Figure A.1.** continued.

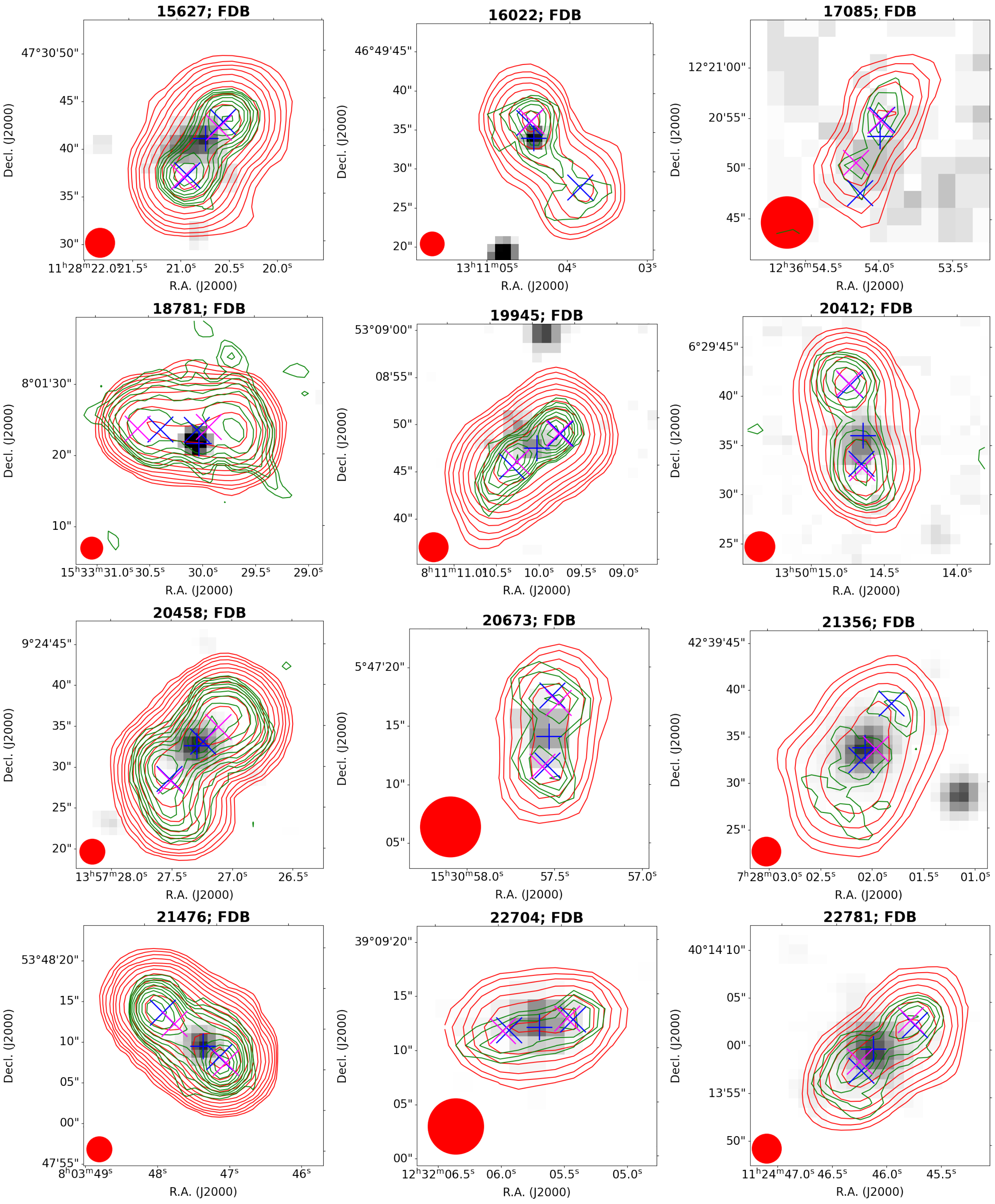

**Figure A.1.** continued.

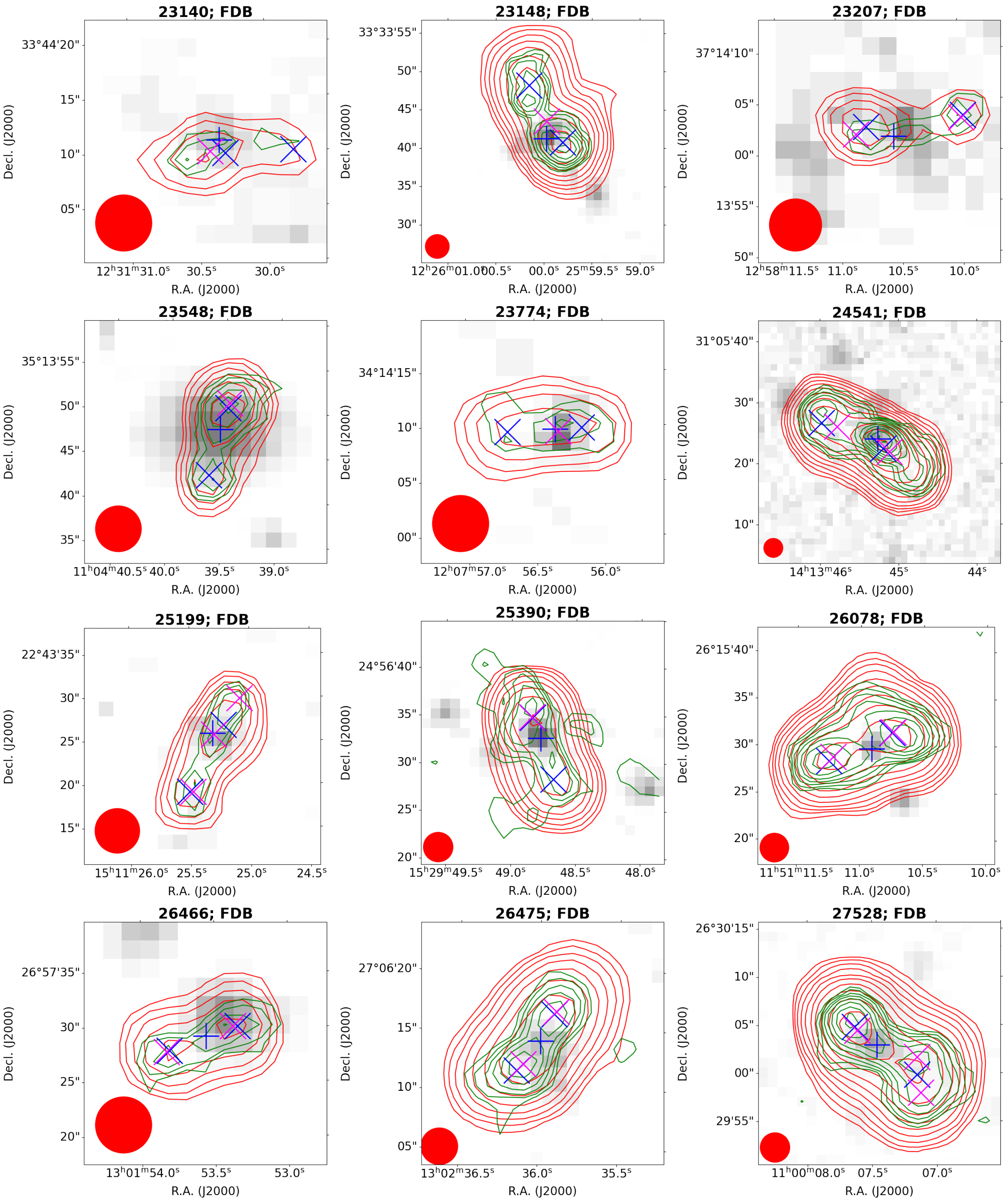

**Figure A.1.** continued.

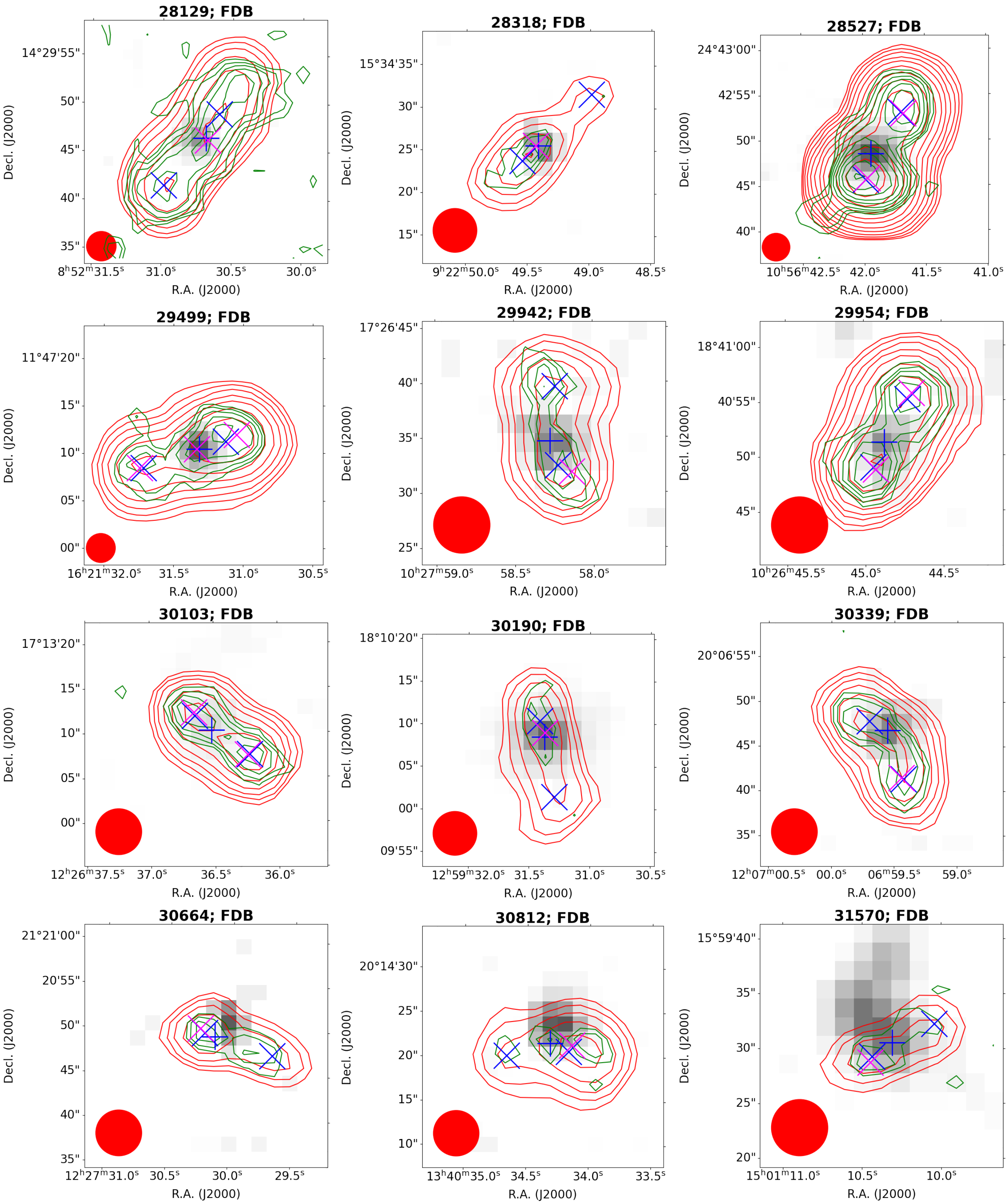

**Figure A.1.** continued.

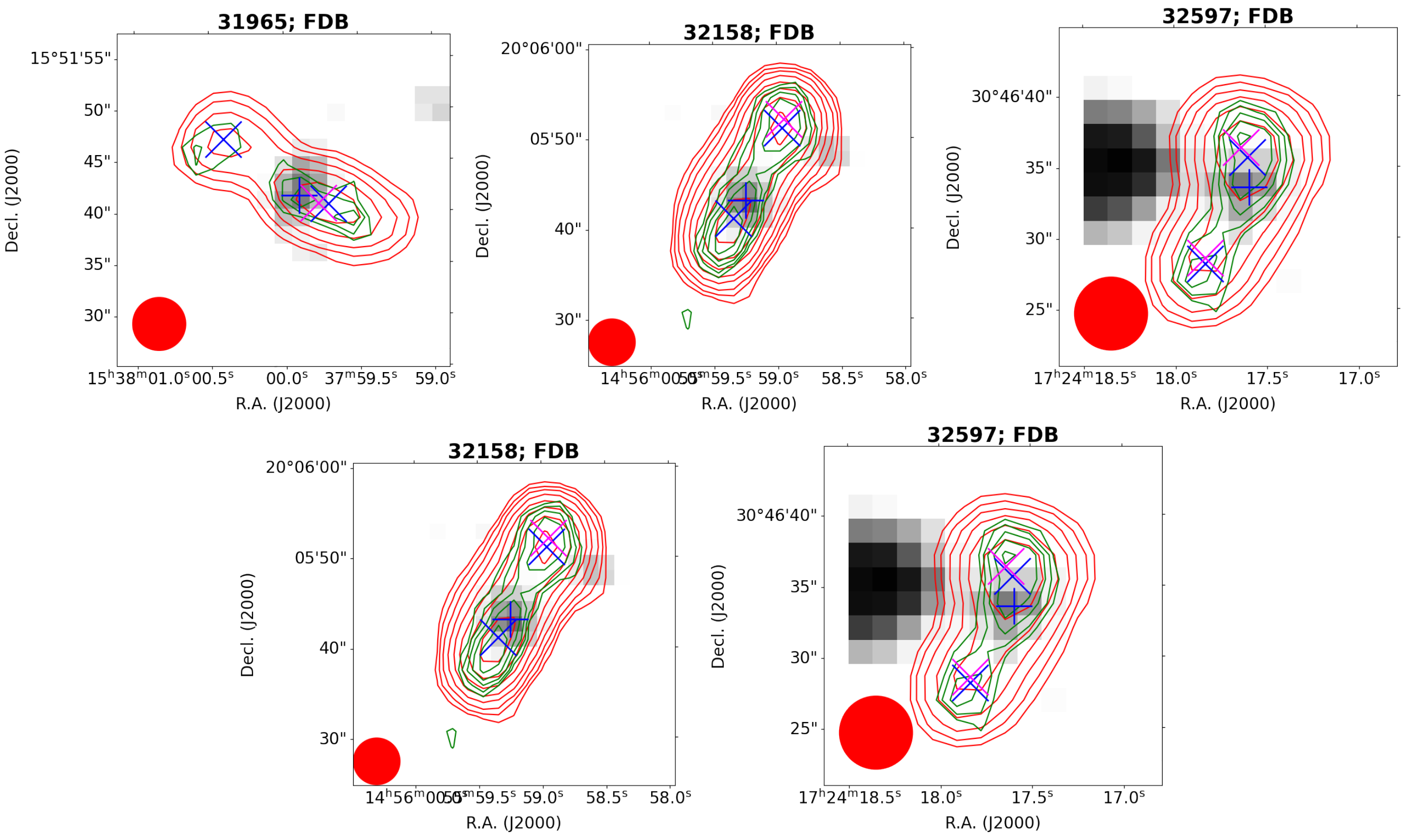

**Figure A.1.** continued.

## B. DISCUSSION ON MALMQUIST BIAS

In order to examine the effect of Malmquist bias on the observed trends between angular size and z (Figure 2c), linear size and z (Figure 2d), flux density and angular size (Figure 2e), and radio power and linear size (Figure 2f) trends, we truncated our sample in total radio luminosity $\geq 3\times10^{24}$ W Hz$^{-1}$, to create a volume-limited sample (B. Efron & V. Petrosian 1992). Figure B.1a shows the total radio power vs. z plot for the full sample, showing the Malmquist bias in our data where intrinsically brighter sources are detected at higher redshifts. The dashed black line shows the limiting luminosity above which all sources are included in our truncated, volume-limited sample, above which the Malmquist bias is less pronounced. This yields a total of 1299 sources (out of 2,002), including 1050 with secure identifications and 249 with possible identifications.

Figure B.1b gives the total angular size vs. z plot. The Spearman rank correlation test gives $\rho = -0.46$ and $p < 10^{-5}$ for a sample with secure identifications only, which changes to $\rho = -0.48$ and $p < 10^{-5}$ for a sample with secure+possible identifications. The $p-$values indicate that the correlation is statistically significant at confidence levels $\geq$99.7% and is moderately strong ($\rho$ ∼-0.5). The index of the power-law fit using linear regression is $\beta \sim -0.8$ (for a sample with secure identifications only), which changes to $\beta \sim -0.9$ (for a sample with secure+possible identifications). The resulting index is close to ∼-1, which is expected for a flat Universe.

Figure B.1c gives the total linear size vs. z plot. The Spearman rank correlation test gives $\rho = -0.03$ and $p < 0.24$ for a sample with secure identifications only, and it changes to $\rho = -0.10$ and $p < 10^{-3}$ for a sample with secure+possible identifications. The small values of $\rho$ <0.1 indicate that the correlation is not statistically significant.

Figure B.1d gives the distribution of 1.4 GHz total flux density with angular size. The Spearman rank correlation test gives the $\rho = -0.69$ and $p < 10^{-5}$, respectively, for a sample with secure identification only, which changes to $\rho = -0.69$ and $p < 10^{-5}$, respectively, for a sample with secure+possible identifications. The $p-$values indicate that the correlation is statistically significant at confidence levels $\geq$99.7% and is moderately strong ($\rho$ ∼0.7). The index of the fitted power-law using linear regression indicates that flux density varies with angular size raised to the power ∼0.7.

Figure B.1e provides the distributions of 1.4 GHz total radio luminosity with total linear size. The Spearman rank correlation test gives $\rho$ and $p-$values of 0.15 and $< 10^{-5}$, respectively, for a sample with secure identifications only, which change to 0.15 and $< 10^{-5}$, respectively, for a sample with secure+possible identifications. The $p-$values indicate that the correlation is statistically significant at confidence levels $\geq$ 99.7% and is weak ($\rho$ ∼0.15). The result of the power-law fitting using linear regression indicates that the total radio power scales as 0.2 power of the projected linear size, flatter than 0.5, obtained using the full sample (Section 4.1.4). This difference can be explained by the Malmquist bias in the full sample in a manner that brighter sources are detected at higher redshifts for a given detection limit of an imaging instrument, and it is the brightest radio AGNs that exhibit the largest sizes (see, Figure 8 of M. J. Hardcastle et al. 2025). This observational bias will steepen the index between radio power and linear size relative to the truncated sample, where the effects of Malmquist bias are less pronounced.

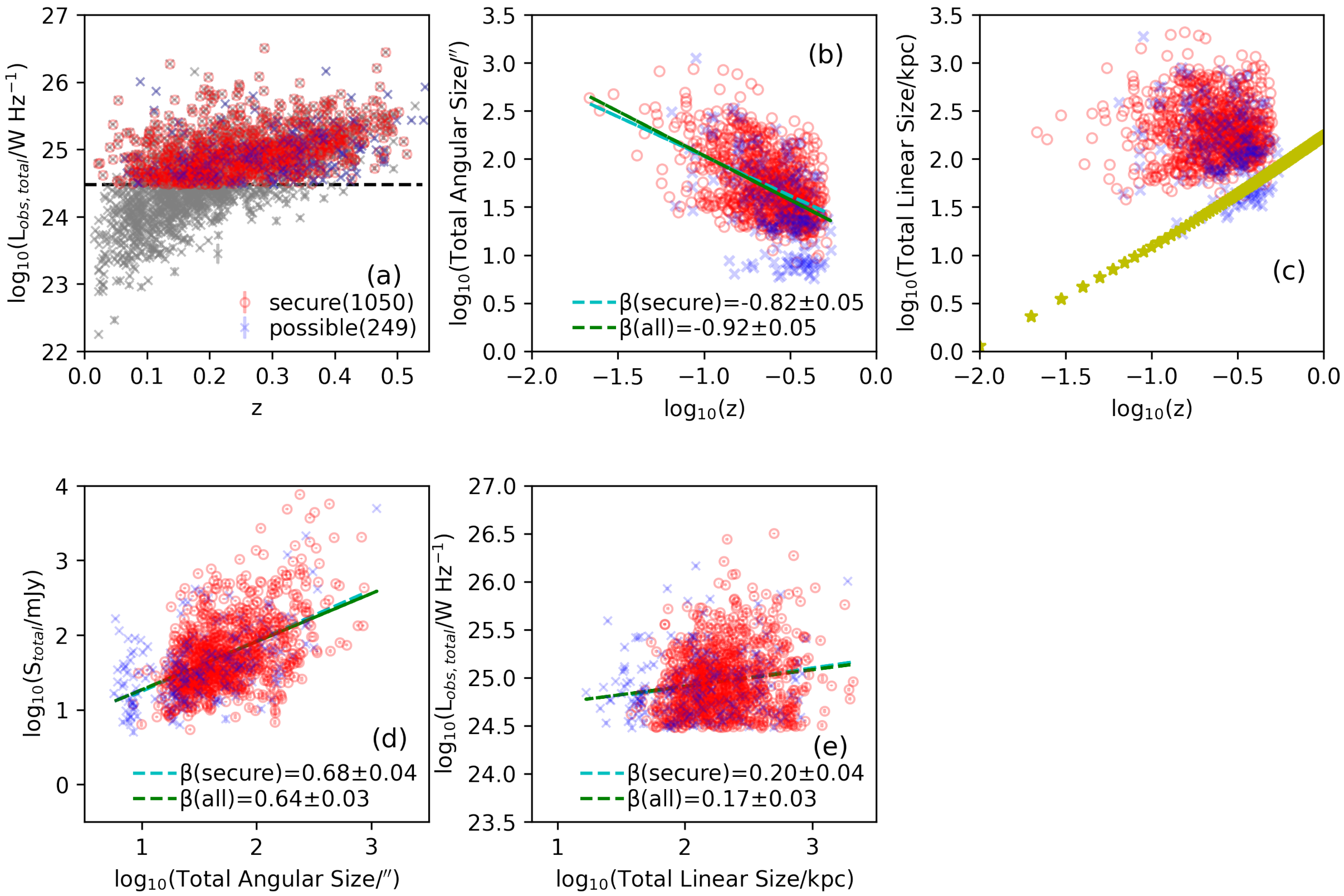

**Figure B.1.** Panel a gives the total radio power vs. z. Gray points include all 2,002 sources, while the truncated sample, consisting of 1,299 sources, is shown in red (secure identification) and blue (possible identification) points. The dashed black line indicates the luminosity limit above which our truncated, volume-limited sample is created, where the effect of Malmquist bias is less pronounced. For this volume-limited sample, the angular size vs. z (panel b), total linear size vs. z (panel c), total flux density vs. angular size (panel c), and total radio power vs. total linear size (panel e) trends are shown. $\beta$(secure), shown by a cyan dashed line, is the index of the power-law fitted between the quantities using the sample of sources with secure morphological identifications only, while $\beta$(all), shown by a green dashed line, is the index using the full sample of sources, i.e., secure+possible morphological identifications while the error on $\beta$ is the standard deviation. Yellow stars in panel c mark the linear sizes corresponding to the beam size of the FIRST survey (5.4″).